\documentclass[twocolumn,aps,pre,amsmath,showpacs,tufte-book]{revtex4-1}
\usepackage{graphicx,amsmath,leftidx,amsfonts}
\usepackage{subfig,array,booktabs,diagbox}

\usepackage[normalem]{ulem}
\usepackage{dcolumn,bm,amssymb,indentfirst}
\usepackage{color,soul}
\setstcolor{red}
\usepackage{caption}
\begin{document}
\author{Luca Banetta$^{1}$, Alessio Zaccone$^{2,1,3}$}
\affiliation{${}^1$Statistical Physics Group, Department of Chemical
Engineering and Biotechnology, University of Cambridge, Cambridge, CB3 0AS, United Kingdom}
\affiliation{${}^2$Department of Physics ``A. Pontremoli", University of Milan, via Celoria 16, 20133 Milan, Italy}
\affiliation{${}^3$Cavendish Laboratory, University of Cambridge, JJ Thomson
Avenue, CB3 0HE Cambridge,
U.K.}
\begin{abstract}
Determining the microstructure of colloidal suspensions under shear flows has been a challenge for theoretical and computational methods due to the singularly-perturbed boundary-layer nature of the problem. Previous approaches have been limited to the case of hard-sphere systems and suffer from various limitations in their applicability. We present a new analytic scheme based on intermediate asymptotics which solves the Smoluchowski diffusion-convection equation including both intermolecular and hydrodynamic interactions. The method is able to recover previous results for the hard-sphere fluid and, for the first time, to predict the radial distribution function (rdf) of attractive fluids such as the Lennard-Jones (LJ) fluid. In particular, a new depletion effect is predicted in the rdf of the LJ fluid under shear. This method can be used for the theoretical modelling and understanding of real fluids subjected to flow, with applications ranging from chemical systems to colloids, rheology, plasmas, and atmospherical science.
\end{abstract}

\title{Radial distribution function of Lennard-Jones fluids in shear flows from intermediate asymptotics}
\maketitle
\section{Introduction}
This work presents a new analytic resolution of the Smoluchowski diffusion equation with shear to analyze the microstructure of strongly sheared colloidal suspensions.
The Smoluchowski equation provides a means to determine the pair distribution function, or radial distribution function (rdf) $g(\textbf{r})$, which gives the probability to find a particle at a certain distance $\textbf{r}$ with respect to a reference (tagged) particle \cite{Hansen,Larsen}.

The rdf is usually influenced by contributions which can be divided into Brownian-induced and shear-induced effects. The first class includes diffusion (Brownian motion) and inter-particle interactions, meanwhile the second class includes the various effects due the flow field.

It is also important to consider the effect of hydrodynamic interactions within the liquid medium: the presence of a second particle will contribute both a shear-induced hydrodynamic interaction \cite{BatchelorGreen} as well as a lubrication contribution \cite{Honig}, if we are under Stokes regime.

The relative importance of shear-induced to Brownian effects is compactly described through the P\'eclet number which, for particles of equal radius $a$, takes on the following form \cite{VandeVen,Morris}:
\begin{equation}
	\text{Pe} =\dfrac{ 6 \pi \eta a^3 \dot{\gamma}}{k_B T}
\label{Péclet_number}
\end{equation}
where $\eta$ is the viscosity of the medium, $k_B$ the Boltzmann constant and $T$ the absolute temperature: if $\text{Pe} \gg 1 $ then the flow field contribution is going to be dominant; as a consequence the Brownian motion is going to be more important if $\text{Pe} \ll 1$.

The analysis of this problem started with the pioneering work of Smoluchowski \cite{Smoluchowski} who evaluated the two asymptotic limits of (i) purely Brownian $\text{Pe} \to 0 $ and (ii) purely convective $\text{Pe} \to \infty$ dynamics, that were used to determine the coagulation rate for the two limits, respectively.

Subsequently, there have been several studies addressing the complex interplay of Brownian-induced and shear-induced contributions to the $g(\textbf{r})$.

Earlier studies adopted various approximation schemes and were addressed to hard-sphere suspensions \cite{Ronis,Hess,Dhont,Blaw} under weak shear flows. Motivation for these earlier studies came from pioneering experiments on hard-sphere colloids where the structure factor distorted by the shear flow was measured with optical techniques~\cite{Ackerson,Clark}. A further motivation comes from rheology: well documented behaviors such as shear thinning of suspensions have their origin in the distortion of the rdf~\cite{Hess,Blaw}, while the rheology of glassy systems under shear can also be described using the Smoluchowski equation with shear as input for the dynamics~\cite{Brader,Cates}. Finally, the microstructure of complex fluids under shear flow is also an essential input for studies which address the dynamics of phase transitions in those systems~\cite{Onuki}.

A parallel line of studies focused on solutions to the Smoluchowski diffusion-convection equation as a means to obtain the coagulation rate of colloids, thus focusing on realistic physico-chemical interactions between the particles~\cite{VandeVen,Feke}.

Several works have addressed the same problem of finding the rdf for haard-sphere suspensions that are subjected to strong shear flows: Batchelor and Green \cite{Batchelor} found the spherically averaged pair distribution function which depends on the radial distance between the particles. It is characterized by an exponential trend which diverges at the contact between the target and the reference particles. Twenty years later, Brady and Morris \cite{MorrisBrady} presented an analytical study on the microstructure of strongly sheared suspensions through perturbative methods and they found a peak in the g($\textbf{r})$ next to the surface of a reference particle whose magnitude is directly proportional to the P\'eclet number. This result has been reconsidered based on Stokesian dynamics simulations \cite{Morris} which found a weaker dependence and highlighted the decrease of the peak with the volume fraction $\phi = 4 \pi n a^3/3$ occupied by the particles, with $n$ the number density. 
Concerning the effect of non-trivial (or non hard-sphere) inter-particle potential on the rdf, which is encountered in most applications, Feke and Schowalter \cite{Feke} considered the case of strongly sheared particles interacting via the Derjaguin-Landau-Verwey-Overbeek (DLVO) potential. Their scheme however relies on a numerical evaluation of the g(\textbf{r}) whose starting point is an approximate far-field analytic resolution which does not go beyond the zeroth order, and is therefore of very limited applicability.

A systematic analytical method to study the influence of shear flow and physico-chemical interactions on the microstructure of sheared suspensions, is still missing.
In the following, we develop a new methodology, based on the rigorous application of intermediate asymptotics, to obtain the rdf as a solution to the Smoluchowski equation with shear flow for the case of Lennard-Jones (LJ) particles. The method provides predictions of the microstructure with non-trivial features due to the interplay between the LJ interaction, the flow field and the Brownian motion. This method can be used in the future to open up new opportunities for elucidating key features of complex fluids, from the rheology of suspensions to problems in atmospheric science. The proposed work goes beyond the current paradigms for hard-sphere systems and presents a solution framework for interacting particles on the example of the 12-6 Lennard-Jones potential which can be easily extended to other important interaction potentials (e.g. Debye-H{\"u}ckel) in the future.

\section{Model}
\begin{figure}[ht]
	\centering
	\includegraphics[width = 0.85 \linewidth]{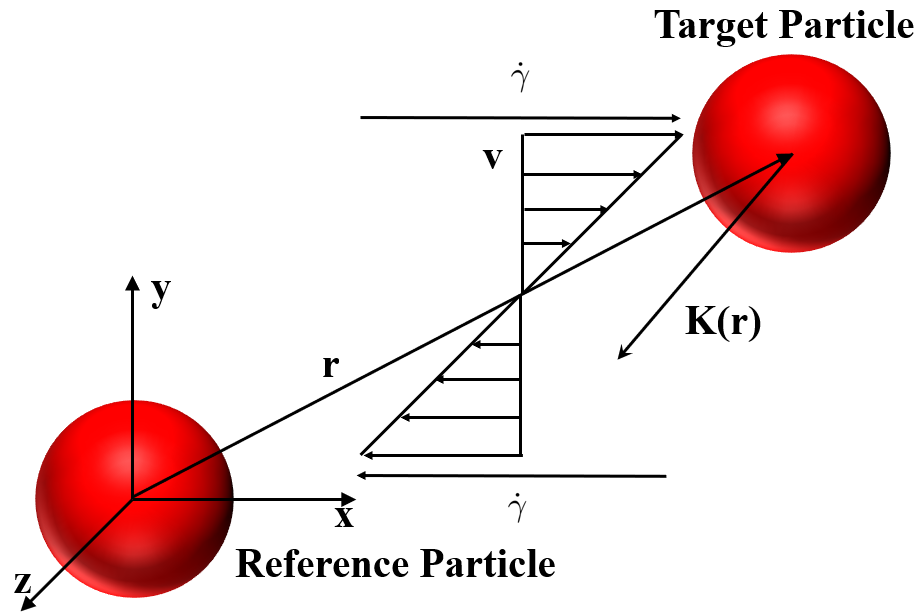}
	\caption{Schematic illustration of a pair of LJ-interacting particles subjected to a simple shear flow where \textbf{v} = (0,$\dot{\gamma}$x,0) in Cartesian coordinates.} .
	\label{fig:physical_system}
\end{figure}

The method is based on finding an analytical solution for the Smoluchowski equation with shear which describes the system depicted in Fig.\ref{fig:physical_system}: we suppose to put a \textit{reference} particle at the centre of the spherical reference frame and a second particle at a certain position $\textbf{r}=(r,\theta,\phi)$.

A simple shear flow is implemented which causes the second particle to have a relative velocity 
$\textbf{v}(\textbf{r})$ with respect to the reference one (or equivalently, this is the velocity of the second particle in the rest frame of the reference particle).
We model the hydrodynamic disturbance of the flow field around the reference particle due to the presence of the target one through the adoption of two functions $A(r)$ and $B(r)$ derived by Batchelor \cite{Batchelor} which influence $\textbf{v}(\textbf{r})$ (see Appendix A for more details). Instead, to describe the effect of lubrication forces we use the widely used parameterized function \cite{Honig,ZacconeNess}:
\begin{equation}
G(h) = \dfrac{6 h^2 + 4 h}{6 h^2 +13 h +2} 
\label{hydrodynamic_function}
\end{equation}
where $h = (r-2 a)/a$ represents the surface-to-surface distance between the particles.

Next we pack the two effects of interparticle potential and shear flow into an external force term $\textbf{K}(\textbf{r})$ acting on the target particle: 
\begin{equation}
    \textbf{K} = -G(r) \nabla{U}(r) + \zeta \textbf{v}(\textbf{r}).
\end{equation}
where $\zeta = 6 \pi \eta a$ is the Stokes friction factor.

\section{Derivation}
Using the functions introduced in the previous section, we can now write the two-body Smoluchowski equation with shear in vector notation:
\begin{equation}
    D \nabla \cdot \biggl( G(r) \nabla g(\textbf{r}) - \dfrac{1}{k_B T} \textbf{K} g(\textbf{r}) \biggr) = 0.
    \label{Smoluchowski_general}
\end{equation}
\textit{D} is the diffusion coefficient, $\textit{k}_B$ the Boltzmann constant and \textit{T} the absolute temperature.\\
Next, we make Eq.(\ref{Smoluchowski_general}) dimensionless through:
\begin{equation}
\begin{cases}
\tilde{\nabla} = \sigma \nabla, \\
\tilde{U} = \dfrac{U}{k_B T}, \\
\end{cases}
\end{equation}
where $\sigma = a$  is the hard-core particle diameter.

The velocity $\textbf{v(r)}$ can be  made dimensionless through:
\begin{equation}
    \textbf{v(r)} =\dot{\gamma} \sigma \tilde{\textbf{v}}(\tilde{\textbf{r}}).
\end{equation}
A more detailed discussion of  $\tilde{\textbf{v}}$ is presented in Appendix A.
Replacing all the previously introduced terms in Eq.(\ref{Smoluchowski_general}) we obtain
\begin{equation}
    \tilde{\nabla} \cdot \biggl[G(\tilde{r}) \tilde{\nabla} g(\tilde{\textbf{r}}) + \left(G(\tilde{r}) \tilde{\nabla} \tilde{U}(\tilde{r}) - \dfrac{ 6 \pi \eta a \sigma^2 \dot{\gamma}}{k_B T} \tilde{\textbf{v}}(\tilde{\textbf{r}}) \right) g(\tilde{\textbf{r}})\biggr] = 0.
    \label{Smoluchowski_dimensionless}
\end{equation}
Expressing the P\'eclet number according to Eq.(\ref{Péclet_number}) we can write:
\begin{equation}
    \tilde{\nabla} \cdot \biggl[ \biggl( G(\tilde{r}) \tilde{\nabla} \tilde{U}(\tilde{r}) - 4 \text{Pe} \tilde{\textbf{v}}(\tilde{\textbf{r}}) \biggr) g(\tilde{\textbf{r}}) + G(\tilde{r}) \tilde{\nabla} g(\tilde{\textbf{r}}) \biggr] = 0.
    \label{Smoluchowski_dimensionless_2}
\end{equation}

Eq.(\ref{Smoluchowski_dimensionless_2}) is to be solved perturbatively. 
A perturbative method approach is based on the introduction of a small \textit{perturbation parameter} $\epsilon$, by definition much smaller than unity, which simplifies the analytical treatment of the partial differential equation (PDE) of interest \cite{BenderOrszag}.

Focusing on situations where the effect of shear flow is substantial, we fix:
\begin{equation}
    \epsilon = \dfrac{1}{\text{Pe}}.
    \label{Perturbationparameter}
\end{equation}
Applying Eq.(\ref{Perturbationparameter}) to Eq.(\ref{Smoluchowski_dimensionless_2}) we obtain:
\begin{equation}
    \tilde{\nabla} \cdot \left[ \epsilon \left(\tilde{\nabla} g(\tilde{\textbf{r}}) + \tilde{\nabla}\tilde{U} g(\tilde{\textbf{r}}) \right)G(\tilde{r}) - 4 \tilde{\textbf{v}}(\tilde{\textbf{r}}) g(\tilde{\textbf{r}}) \right] = 0.
    \label{Smoluchowski_epsilon}
\end{equation}
Starting from Eq.(\ref{Smoluchowski_epsilon}) we apply the linearity of the divergence operators obtaining:
\begin{multline}
    \epsilon \biggl( G(\tilde{r}) \tilde{\nabla}^2 g(\tilde{\textbf{r}}) + \tilde{\nabla} G(\tilde{r}) \cdot \tilde{\nabla} g(\tilde{\textbf{r}}) + G(\tilde{r}) \tilde{\nabla} \tilde{U} \cdot \tilde{\nabla} g(\tilde{\textbf{r}}) +\\+ g(\tilde{\textbf{r}}) \tilde{\nabla} G(\tilde{r}) \cdot \tilde{\nabla} \tilde{U} + G(\tilde{r}) \tilde{\nabla}^2 \tilde{U} g(\tilde{\textbf{r}}) \biggr) + \\ - 4 \biggl( \tilde{\textbf{v}} \cdot \tilde{\nabla} g(\tilde{\textbf{r}}) + g(\tilde{\textbf{r}}) \tilde{\nabla}\cdot\tilde{\textbf{v}} \biggr) = 0.
    \label{Smoluchowski_epsilon_1}
\end{multline}
To be as general as possible, we do not neglect the compressibility of the fluid throughout the following manipulations, considering the divergence of the velocity field to be not null.

Next, we introduce a useful approximation that was proposed in \cite{ZacconePRE2009,NazockdastMorris} in order to make the 3D problem analytically solvable. The approximation consists of the application of an angular average, denoted as $\langle \cdots \rangle$, over a solid angle to Eq.(\ref{Smoluchowski_epsilon_1}). 
\begin{multline}
    \epsilon \biggl( G(\tilde{r}) \tilde{\nabla}^2 \langle g(\tilde{\textbf{r}}) \rangle + \tilde{\nabla} G(\tilde{r}) \cdot \tilde{\nabla}\langle g(\tilde{\textbf{r}}) \rangle + G(\tilde{r}) \tilde{\nabla} \tilde{U} \cdot \tilde{\nabla} \langle g(\tilde{\textbf{r}}) \rangle +\\+ \langle g(\tilde{\textbf{r}}) \rangle \tilde{\nabla} G(\tilde{r}) \cdot \tilde{\nabla} \tilde{U} + G(\tilde{r}) \tilde{\nabla}^2 \tilde{U} \langle g(\tilde{\textbf{r}}) \rangle \biggr) + \\ - 4 \biggl( \langle \tilde{\textbf{v}} \cdot \tilde{\nabla} g(\tilde{\textbf{r}}) \rangle + \langle g(\tilde{\textbf{r}}) \tilde{\nabla}\cdot\tilde{\textbf{v}} \rangle \biggr) = 0.
\end{multline}
The result is the following spherically-averaged solution $g(\tilde{r})$ which depends on the radial coordinate only:
\begin{multline}
\epsilon \biggl( G(\tilde{r}) \tilde{\nabla}^2 g(\tilde{r}) + \tilde{\nabla} G(\tilde{r}) \cdot \tilde{\nabla} g(\tilde{r}) + G(\tilde{r}) \tilde{\nabla} \tilde{U} \cdot \tilde{\nabla} g(\tilde{r}) +\\+ g(\tilde{r}) \tilde{\nabla} G(\tilde{r}) \cdot \tilde{\nabla} \tilde{U} + G(\tilde{r}) \tilde{\nabla}^2 \tilde{U} g(\tilde{r}) \biggr) + \\ - 4 \biggl( \langle \tilde{\textbf{v}} \cdot \tilde{\nabla} g(\tilde{\textbf{r}}) \rangle + \langle g(\tilde{\textbf{r}}) \tilde{\nabla}\cdot\tilde{\textbf{v}} \rangle \biggr) = 0.
\end{multline}
Moreover, we use a \textit{decoupling approximation}: we suppose that the velocity and the pair correlation function are weakly correlated, so that:
\begin{equation}
    \langle \tilde{\textbf{v}} \cdot \tilde{\nabla} g(\tilde{\textbf{r}}) \rangle + \langle g(\tilde{\textbf{r}}) \tilde{\nabla}\cdot \tilde{\textbf{v}} \rangle \approx  \langle \tilde{\textbf{v}} \rangle \cdot \tilde{\nabla} g(\tilde{r}) + g(\tilde{r}) \langle \tilde{\nabla} \cdot \textbf{v} \rangle .
\end{equation}

A general flow field can be separated into downstream and an upstream regions: in the former regions the particles approach each other (\textit{compressing} sectors of solid angle), so the relative velocity between the particles is negative; instead, in the upstream regions (\textit{extensional} sectors), the particles move away from each other, leading to a positive radial velocity.
Within this methodology, the actual relative velocity and the flow field divergence are replaced with their angular averaged values within downstream and upstream regions.

The angular averaging is necessary in order to reduce the original PDE (which is solvable only numerically, and even then poses quite some computational challenges) to an ODE which is analytically solvable. The price to pay for having analytical solutions is that it is not possible to produce deformed contour plots to highlight the angle-dependent rdfs. 

Now, we will consider two coefficients which are the results of the average procedure,: $\alpha_c$ for the downstream and $\alpha_e$ for the upstream zone, which are explicitly introduced and defined in Appendix A. The two coefficients define the influence of the angular coordinates on the radial relative velocity and the flow field divergence as shown in Eq.(\ref{relative_velocity_and_divergence}). Thus for the compressive quadrants we have:
\begin{equation}
\begin{cases}
\langle \tilde{\textbf{v}} \rangle = \alpha_c (1-A(\tilde{r})) \tilde{r}, \\
\langle \tilde{\nabla} \cdot \textbf{v} \rangle = \alpha_c \biggl( 3(B(\tilde{r})-A(\tilde{r})) - \tilde{r} \dfrac{d A(\tilde{r})}{d \tilde{r}}\biggr)
\end{cases}
\label{relative_velocity_and_divergence}
\end{equation}
and analogous expressions are found for the extensive quadrants.
Next, we arrive at the following expression:
\begin{multline}
\epsilon \biggl( G(\tilde{r}) \tilde{\nabla}^2 g(\tilde{r}) + \tilde{\nabla} G(\tilde{r}) \cdot \tilde{\nabla} g(\tilde{r}) + G(\tilde{r}) \tilde{\nabla} \tilde{U} \cdot \tilde{\nabla} g(\tilde{r}) +\\+ g(\tilde{r}) \tilde{\nabla} G(\tilde{r}) \cdot \tilde{\nabla} \tilde{U} + G(\tilde{r}) \tilde{\nabla}^2 \tilde{U} g(\tilde{r}) \biggr) + \\ - 4 \biggl( \langle \tilde{\textbf{v}} \rangle \cdot \tilde{\nabla} g(\tilde{r}) +  g(\tilde{r}) \langle \tilde{\nabla}\cdot\tilde{\textbf{v}} \rangle \biggr) = 0.
\label{Smoluchowski_averaged_3D}
\end{multline}
Then, we isolate the dependence on the radial coordinate only, obtaining:
\begin{multline}
    \epsilon \biggl[ \dfrac{G(\tilde{r})}{\tilde{r}^2} \dfrac{\text{d}}{\text{d} \tilde{r}} \biggl( \tilde{r}^2 \dfrac{\text{d} g(\tilde{r})}{\text{d} \tilde{r}} \biggr) + \dfrac{\text{d} G}{\text{d} \tilde{r}} \dfrac{\text{d} g(\tilde{r})}{\text{d} \tilde{r}} + G(\tilde{r}) \dfrac{\text{d} \tilde{U}}{\text{d} \tilde{r}} \dfrac{\text{d} g(\tilde{r})}{\text{d} \tilde{r}} + \\ + g(\tilde{r}) \dfrac{\text{d} G}{\text{d} \tilde{r}} \dfrac{\text{d} \tilde{U}}{\text{d} \tilde{r}} + \dfrac{G(\tilde{r})}{\tilde{r}^2} \dfrac{\text{d}}{\text{d} \tilde{r}} \biggl( \tilde{r}^2 \dfrac{\text{d} \tilde{U}}{\text{d} \tilde{r}} \biggr) g(\tilde{r}) \biggr]+ \\ - 4 \biggl( \langle \tilde{\textbf{v}} \rangle \dfrac{\text{d} g(\tilde{r})}{\text{d}\tilde{r}} + g(\tilde{r}) \langle \tilde{\nabla} \cdot \textbf{v} \rangle \biggr) = 0.
\end{multline}
Finally, we put the equation in the following final form which is the most convenient for the perturbative treatment:
\begin{multline}
    \epsilon \biggl[ G(\tilde{r}) \biggl( \dfrac{\text{d}^2 g}{\text{d} \tilde{r}^2}  + \dfrac{2}{\tilde{r}} \dfrac{\text{d} g}{\text{d}\tilde{r}}\biggr) + \dfrac{\text{d} G}{\text{d}\tilde{r}} \dfrac{\text{d} g}{\text{d}\tilde{r}}+ g \dfrac{\text{d} \tilde{U}}{\text{d} \tilde{r}} \dfrac{\text{d} G}{\text{d} \tilde{r}}+ \\ + G \dfrac{\text{d} \tilde{U}}{\text{d} \tilde{r}} \dfrac{\text{d} g}{\text{d} \tilde{r}} + G \biggl( \dfrac{2}{\tilde{r}} \dfrac{\text{d}\tilde{U}}{\text{d} \tilde{r}} + \dfrac{\text{d}^2 \tilde{U}}{\text{d} \tilde{r}^2} \biggr)g(\tilde{r}) \biggr] + \\ -  4 \biggl( \langle \tilde{\textbf{v}} \rangle \dfrac{\text{d} g}{\text{d}\tilde{r}} + g \langle \tilde{\nabla} \cdot \textbf{v} \rangle \biggr) = 0.
    \label{Smoluchowski_epsilon_final}
 \end{multline}
Since Eq.(\ref{Smoluchowski_epsilon_final}) is a second order differential equation, we need two boundary conditions (BCs). The first one is the usual the far-field BC:
\begin{equation}
g(\tilde{r} \to \infty) = 1.
\label{BC_1}
\end{equation}
The second BC expresses the fact that the radial flux is null when the two particles are in direct contact:
\begin{equation}
G(\tilde{r}_c) \biggl( \dfrac{\text{d}g}{\text{d}\tilde{r}} \biggr)(\tilde{r}_c) + \biggl( G(\tilde{r}_c) \dfrac{\text{d} \tilde{U}}{\text{d}\tilde{r}} - 4 \text{Pe} \langle \tilde{\textbf{v}} \rangle \biggr) g(\tilde{r}_c) = 0,
\label{BC_2}
\end{equation}
where $\tilde{r}_c$ is a value of radial distance sufficiently close to the reference particle. In our calculations we will take $\tilde{r}_c = 1 + 5 \times 10^{-5}$.

From inspection of Eq.(\ref{Smoluchowski_epsilon_final}) it can immediately be seen that the perturbation parameter is linked with the highest order derivative of the ordinary differential equation (ODE). This means that we are dealing with a singular perturbation problem and it must be solved by the application of the boundary layer theory \cite{BenderOrszag}.\\
The approach consists of the evaluation of two different power series related to two different regions of the domain: the outer layer (in this case farther away the reference particle), where the solution is slowly changing with $\tilde{r}$, and the inner region (closer to the reference particle), usually called boundary layer, where the solution is steeply and very rapidly changing with the radial coordinate.

\subsection{Outer solution}
We write the outer solution as a power series in the small parameter $\epsilon$:
\begin{equation}
    g^{\text{out}} = g_{0}^{\text{out}} + \epsilon  g_{1}^{\text{out}} + \epsilon^2  g_{2}^{\text{out}} + \dots .
    \label{outersolution}
\end{equation}
We will carry out our derivation up to first order in the perturbative series.

The initial step is the assumption of the solution to be characterized by low values of its derivatives. As a consequence, they become negligible if they are multiplied by a small parameter such as $\epsilon$; this concept can be implemented by imposing $\epsilon=0$ in Eq.(\ref{Smoluchowski_epsilon_final}).

It should be emphasized that for the outer solution we obtain a first order boundary value problem, so we are forced to impose only one of the two BCs. Since we know that the boundary layer is close to $\tilde{r}_c$ we choose Eq.(\ref{BC_1}), thus ending up with 
\begin{equation}
\begin{cases}
\langle \tilde{\textbf{v}} \rangle \dfrac{\text{d} g_0^{\text{out}}(\tilde{r})}{\text{d} r} + g_0^{\text{out}}(\tilde{r}) \langle \tilde{\nabla} \cdot \textbf{v} \rangle = 0.\\
g_{0}^{\text{out}}(\tilde{r} \to \infty) = 1
\label{leading_order_solution}
\end{cases}
\end{equation}
which leads to the following solution (derived with full details in Appendix B):
\begin{equation}
    g_{0}^{\text{out}} = \dfrac{1}{1-A(\tilde{r})} \exp\biggl[{\int_{\tilde{r}}^{\infty}\biggl(\dfrac{3(B-A)}{\tilde{r}(1-A)}\biggr)}d \tilde{r} \biggr].
    \label{leading_outer_term}
\end{equation}
Equation (\ref{leading_outer_term}) is formally identical to the well-known solution proposed by Batchelor and Green~\cite{BatchelorGreen}, an evidence of the good reliability of the method.\\
Now, we will evaluate the first order term $g_1^{\text{out}}$; starting from Eq.(\ref{outersolution}) we need to evaluate the first and second derivative of $g^{\text{out}}$:
\begin{equation}
\begin{cases}
    \dfrac{\text{d} g^{\text{out}}}{\text{d} \tilde{r}} = \dfrac{\text{d} g_{0}^{\text{out}}}{\text{d} \tilde{r}} + \epsilon \dfrac{\text{d} g_{1}^{\text{out}}}{\text{d} \tilde{r}} + \dots \\ \\
    \dfrac{\text{d}^2 g^{\text{out}}}{\text{d} \tilde{r}^2} = \dfrac{\text{d}^2 g_{0}^{\text{out}}}{\text{d} \tilde{r}^2} + \epsilon \dfrac{\text{d}^2 g_{1}^{\text{out}}}{\text{d} \tilde{r}^2} + \dots 
\end{cases}
    \label{outer_gradient}
\end{equation}
Then we replace Eq.(\ref{outer_gradient}) and Eq.(\ref{outersolution}) in Eq.(\ref{Smoluchowski_epsilon_final}) and we group all the terms linked with the same power of the perturbation parameter. Since the perturbation parameter can never be null, the only way to find the n-th order coefficient of Eq.(\ref{outersolution}) is to impose the coefficient related to the n-th power of $\epsilon$ to be zero.

In particular, to find $g_1^{\text{out}}$, in Eq.(\ref{firstorderouterequation}) we isolate the coefficient which multiplies the first power of $\epsilon$:
\begin{multline}
\begin{cases}
\biggl[  G \biggl( \dfrac{\text{d}^2  g_{0}^{\text{out}}}{\text{d} \tilde{r}^2}  + \dfrac{2}{\tilde{r}} \dfrac{\text{d}  g_{0}^{\text{out}}}{\text{d}\tilde{r}}\biggr) + \dfrac{\text{d} G}{\text{d}\tilde{r}} \dfrac{\text{d}  g_{0}^{\text{out}}}{\text{d}\tilde{r}}+ g_{0}^{\text{out}} \dfrac{\text{d} \tilde{U}}{\text{d} \tilde{r}} \dfrac{\text{d} G}{\text{d} \tilde{r}}+ \\ + G \dfrac{\text{d} \tilde{U}}{\text{d} \tilde{r}} \dfrac{\text{d}  g_{0}^{\text{out}}}{\text{d} \tilde{r}} + G \biggl( \dfrac{2}{\tilde{r}} \dfrac{\text{d}\tilde{U}}{\text{d} \tilde{r}} + \dfrac{\text{d}^2 \tilde{U}}{\text{d} \tilde{r}^2} \biggr) g_{0}^{\text{out}} \biggr] + \\ -  4 \biggl( \langle \tilde{\textbf{v}} \rangle \dfrac{\text{d} g_1^{\text{out}}}{\text{d} \tilde{r}} + g_{1}^{\text{out}} \langle \tilde{\nabla} \cdot \textbf{v} \rangle \biggr) = 0,\\ \\
g_{1}^{\text{out}}(\tilde{r} \to \infty) = 0,
    \label{firstorderouterequation}
\end{cases}
\end{multline}
The boundary condition in Eq.(\ref{firstorderouterequation}) represents the independence of far-field boundary condition from the P\'eclet number.

It is possible to solve Eq.(\ref{firstorderouterequation}) analytically following the steps reported in Appendix B leading to:
\begin{multline}
    g_{1}^{\text{out}} = - g_0^{\text{out}} \int_{\tilde{r}}^{\infty} \dfrac{1}{4 \langle \tilde{\textbf{v}} \rangle} \biggl \{ G \biggl[ Y^2 + \dfrac{\text{d} Y}{\text{d} \tilde{r}} + \biggl( \dfrac{2}{\tilde{r}} + \dfrac{\text{d} \tilde{U}}{\text{d} \tilde{r}}\biggr) Y(\tilde{r}) + \\ + \dfrac{\text{d}^2\tilde{U}}{\text{d} \tilde{r}^2} + \dfrac{2}{\tilde{r}}\dfrac{\text{d} \tilde{U}}{\text{d} \tilde{r}} \biggr] + \dfrac{\text{d}G}{\text{d}\tilde{r}}\biggl( Y + \dfrac{\text{d} \tilde{U}}{\text{d}\tilde{r}} \biggr) \biggr\} \text{d} \tilde{r},
    \label{firstorderouterfinal}
\end{multline}
where $Y(\tilde{r})$ is a function introduced in order to simplify the structure of the first order term; its full expression has been explicitly defined in Appendix B.
Finally, we can obtain the first order angular-averaged outer solution:
\begin{multline}
g^{\text{out}} = \biggl[ 1 - \epsilon \biggl ( \int_{\tilde{r}}^{\infty} \dfrac{1}{4 \langle \tilde{\textbf{v}} \rangle} \biggl \{ G \biggl[ Y^2 + \dfrac{\text{d} Y}{\text{d} \tilde{r}} + \biggl( \dfrac{2}{\tilde{r}} + \dfrac{\text{d} \tilde{U}}{\text{d} \tilde{r}}\biggr) Y(\tilde{r}) + \\ + \dfrac{\text{d}^2\tilde{U}}{\text{d} \tilde{r}^2} + \dfrac{2}{\tilde{r}}\dfrac{\text{d} \tilde{U}}{\text{d} \tilde{r}} \biggr] + \dfrac{\text{d}G}{\text{d}\tilde{r}}\biggl( Y + \dfrac{\text{d} \tilde{U}}{\text{d}\tilde{r}} \biggr) \biggr\} \text{d} \tilde{r}\biggr) \biggr] g_0^{\text{out}} + O(\epsilon^2).
\label{outersolutionfinal}
\end{multline}

\subsection{Inner Solution}
We now focus on the small section of the domain where the solution varies dramatically with respect to variations in $\tilde{r}$.
The first step is the application of change of variable called $\textit{inner transformation}$. Since it is known that the region where the rdf varies the most is close to the reference particle, we can define the inner coordinate $\xi$ as:
\begin{equation}
\xi = \dfrac{\tilde{r}-\tilde{r}_c}{\delta(\epsilon)}
    \label{innervariable}
\end{equation}
where $\delta(\epsilon)$ is the order of magnitude of the width of the boundary layer.

Before turning to the evaluation of the inner solution $g^{\text{in}}$ we must find the relationship between $\delta(\epsilon)$ and the perturbation parameter $\epsilon$ itself; this procedure is called $\textit{dominant balancing}$ \cite{BenderOrszag,Hinch}.

At first, we need to apply the inner transformation to Eq.(\ref{Smoluchowski_epsilon_final}), giving
\begin{multline}
        \epsilon \biggl[ G(\xi)\biggl( \dfrac{\text{d}^2 g^{\text{in}}(\xi)/\text{d} \xi^2}{\delta^2(\epsilon)} + \dfrac{2}{(\xi \delta(\epsilon) + \tilde{r}_c)}\dfrac{\text{d} g^{\text{in}}/\text{d}\xi}{\delta(\epsilon)} \biggr)+\\+\dfrac{(\text{d}G/\text{d}\xi) (\text{d} g^{\text{in}}/\text{d}\xi)}{\delta(\epsilon)^2} + G(\xi)\biggl( \dfrac{\text{d}^2 \tilde{U}/\text{d}\xi^2}{\delta(\epsilon)^2}+ \\  + \dfrac{2}{(\xi \delta(\epsilon) +  \tilde{r}_c)} \dfrac{\text{d} \tilde{U} /\text{d} \xi}{\delta(\epsilon)} \biggr) +G(\xi) \dfrac{(\text{d}\tilde{U}/ \text{d} \xi) (\text{d}g^{\text{in}}/\text{d}\xi)}{\delta(\epsilon)^2}+ \\ + g^{\text{in}} \dfrac{(\text{d}\tilde{U}/\text{d}\xi) (\text{d}G/\text{d}\xi)}{\delta(\epsilon)^2}\biggr] + \\  - 4 \biggl( \dfrac{\langle \tilde{\textbf{v}}(\xi)\rangle}{\delta(\epsilon)}  \text{d} g^{\text{in}}/ \text{d} \xi + g^{\text{in}} \langle \tilde{\nabla} \cdot \textbf{v}(\xi) \rangle \biggr) = 0.
    \label{Smoluchowski_epsilon_inner}
\end{multline}
The crucial point of the dominant balancing is to find the asymptotic behavior of all the functions in the above ODE with respect to $\delta(\epsilon)$ and $\epsilon$.
For this reason we need to find a set of ``gauge functions" representing the asymptotic trend of every term in Eq.(\ref{Smoluchowski_epsilon_inner}) as $\epsilon \to 0$ \cite{VanDyke}; all the mathematical steps are reported in Appendix C and it has been found that
\begin{equation}
    \begin{cases}
    
    \dfrac{\text{d} \tilde{U}}{\text{d} \xi} = O(\delta(\epsilon)) = \delta(\epsilon) W(\xi) \\ \\
    \dfrac{\text{d}^2 \tilde{U}}{\text{d} \xi^2} =  O(\delta(\epsilon)^2) = \delta(\epsilon)^2 X(\xi) \\ \\
    \langle \tilde{\nabla} \cdot \textbf{v}(\xi) \rangle = O(1) \\ \\
    \dfrac{\text{d}G}{\text{d}\xi} = O(\delta(\epsilon)) = \delta(\epsilon) G_r
     
    \end{cases}
    \label{interaction_derivatives}
\end{equation}
where $W(\xi)$, $X(\xi)$, $\langle \tilde{\nabla} \cdot \textbf{v}(\xi) \rangle$ and $G_r$ are functions supposed to be bounded as $\epsilon \to 0$, so they will be considered as $O(1)$. It is important to highlight that the terms related to the inter-particle potential have been evaluated considering a 12-6 Lennard-Jones potential.\\

Upon replacing all the above expressions in Eq.(\ref{Smoluchowski_epsilon_inner}) and rearranging we obtain:
\begin{multline}
  \dfrac{\epsilon}{\delta(\epsilon)} \biggl( G(\xi) \dfrac{\text{d}^2 g^{\text{in}}}{ \text{d} \xi^2} \biggr) + \epsilon \biggl( \dfrac{2 G(\xi)}{(\xi \delta(\epsilon) + \tilde{r}_c)} + G(\xi) W(\xi) + G_r(\xi) \biggr)\\ \times \dfrac{\text{d} g^{\text{in}}}{\text{d}\xi} + \epsilon \delta(\epsilon) \biggl[G \biggl( \dfrac{2}{(\xi \delta(\epsilon) + \tilde{r}_c)}  W(\xi) +  X(\xi) \biggr) + W(\xi) G_r(\xi) \biggr] \\ \times g^{\text{in}} - 4 \biggl( \langle \tilde{\textbf{v}} \rangle \dfrac{\text{d} g^{\text{in}}}{\text{d} \xi} + \delta(\epsilon) g^{\text{in}} \langle \tilde{\nabla}_{\xi} \cdot \tilde{\textbf{v}} \rangle \biggr) = 0.
    \label{Smoluchowski_epsilon_inner_1}  
\end{multline}
The gauge functions representing the orders of magnitude of the terms in Eq.(\ref{Smoluchowski_epsilon_inner_1}) are listed in Table \ref{coefficients_dominat_balancing}.

\begin{table}
	\centering
	\begin{tabular}{|c|c|c|c|c|}
		\hline      
		\textbf{(i)} & \textbf{(ii)} & \textbf{(iii)} & \textbf{(iv)} & \textbf{(v)} \\
		\hline
	O($\epsilon/\delta(\epsilon))$ & O($\epsilon$) & O($\epsilon \delta(\epsilon)$) & O(1) & O($\delta(\epsilon)$)\\
		\hline
	\end{tabular}
	\caption{coefficients appearing in Eq.(\ref{Smoluchowski_epsilon_inner_1})} as they appear in the equation from left to right.
	\label{coefficients_dominat_balancing}
\end{table}
 
The coefficient (iv) multiplying the velocity has to be $O(1)$ because the relative velocity between the particles can never be null.
The aim of this procedure is to find the pair of coefficients of Eq.(\ref{Smoluchowski_epsilon_inner_1}) which counts the most as $\epsilon \to 0$, through a "trial and error" procedure; immediately we can notice that, since both $\epsilon$ and $\delta(\epsilon)$ are much less then unity, (iii) is going to be for sure less dominant than the other terms, so it will be discarded.

Let us now suppose that the pair of terms that dominate are (ii) and (iv): this choice is unreasonable since we have assumed at the beginning of the derivation that $\epsilon \ll 1$. 
Next, if we assume that (i) and (ii) are dominant as $\epsilon \to 0$, this means that $\delta(\epsilon) = O(1)$. In this case (i), (ii), (iv) and (v) become respectively $O(\epsilon$), O($\epsilon$), O(1) and O($\epsilon$): it is evident that the second last term, considered to be negligible, is actually the most important one as $\epsilon \to 0$, so even this hypothesis is unreasonable.

Finally, we suppose that (i) and (iv) are the most dominant terms, which leads to $\delta(\epsilon) = O(\epsilon)$. This final assumption is the correct one because (i), (ii),(iv) and (v) become $O(1)$, $O(\epsilon)$, $O(1)$ and O($\epsilon$) respectively; in this case the excluded terms are negligible compared to the other two as $\epsilon \to 0$; for this reason we will consider the following expression for the boundary layer:
\begin{equation}
    \delta(\epsilon) = O(\epsilon) = O(\text{Pe}^{-1}).
    \label{boundarylayer_gauge_function}
\end{equation}
Eq.(\ref{boundarylayer_gauge_function}) provides an estimate of the width of the boundary layer which is in agreement with the literature \cite{Feke}, but it has been derived here in a rigorous way for the first time.

Based upon the previous results, we can write the differential equation for the evaluation of $g^{\text{in}}$ in its final form as:
\begin{multline}
  G(\xi) \dfrac{\text{d}^2 g^{\text{in}}}{\text{d} \xi^2} + \epsilon \biggl[ \biggl(\dfrac{2 G(\xi)}{(\xi \epsilon + \tilde{r}_c)}  + G(\xi) W(\xi)  + G_r(\xi) \biggr) \times \\ \times \dfrac{\text{d} g^{\text{in}}(\xi)}{\text{d} \xi} - 2 \langle \tilde{\nabla}_{\xi} \cdot \textbf{v}(\xi) \rangle  g^{\text{in}}(\xi) \biggr]+ \epsilon^2 g^{\text{in}}(\xi) \times \\ \times \biggl( X(\xi)G(\xi) + 2 \dfrac{G(\xi)W(\xi)}{(\xi \epsilon + \tilde{r}_c)} + G_r(\xi)W(\xi) \biggr) + \\ -  4 \langle \tilde{\textbf{v}}(\xi) \rangle  \dfrac{\text{d} g^{\text{in}}(\xi)}{\text{d} \xi} = 0.
\label{inner_equation_final}
\end{multline}
Since we want the inner solution to be a series based on the powers of a small parameter, we can adopt the result previously obtained in Eq.(\ref{boundarylayer_gauge_function}) to describe the structure of $g^{\text{in}}$ as :
\begin{equation}
g^{\text{in}} = g_{0}^{\text{in}} + \epsilon g_{1}^{\text{in}} + \epsilon^2 g_{2}^{\text{in}} + \dots
\label{actualinnersolution}
\end{equation}
Now it is possible to evaluate the leading order term of Eq.(\ref{actualinnersolution}); under the same hypothesis made for the outer solution we impose $\epsilon \to 0$ in Eq.(\ref{inner_equation_final}) ending up with the following boundary value problem:
\begin{equation}
\dfrac{\text{d}^2 g_{0}^{\text{in}}}{\text{d} \xi^2} - 4 \dfrac{\langle \tilde{\textbf{v}}(\epsilon = 0) \rangle}{G(\epsilon = 0)} \dfrac{\text{d} g_{0}^{\text{in}}}{\text{d} \xi} = 0.
\label{leadinginnerequation}
\end{equation}
The solution of Eq.(\ref{leadinginnerequation}) is straightforward and we find:
\begin{equation}
    g_{0}^{\text{in}} = C_1 + C_0 \int_0^{\xi} \exp \biggl[ \biggl(\int_0^{\xi} 4 \dfrac{\langle \tilde{\textbf{v}}(\epsilon = 0) \rangle}{G(\epsilon = 0)} d\xi \biggr) \biggr] d\xi
\end{equation}
where $C_0$ and $C_1$ are integration constants to be evaluated later.\\ 

Considering the behavior of $g_{1}^{\text{in}}$, we need to replace the first and the second order derivative of the inner solution, defined as
\begin{equation}
    \begin{cases}
    \dfrac{\text{d} g^{\text{in}}}{\text{d} \xi}   = \dfrac{\text{d} g_{0}^{\text{in}}}{\text{d} \xi} + \epsilon \dfrac{\text{d} g_{1}^{\text{in}}}{\text{d} \xi}  + \epsilon^2 \dfrac{\text{d}g_{2}^{\text{in}}}{\text{d} \xi} + \dots. \\ \\
\dfrac{\text{d}^2 g^{\text{in}}}{\text{d} \xi^2}   = \dfrac{\text{d}^2 g_{0}^{\text{in}}}{\text{d} \xi^2} + \epsilon \dfrac{\text{d}^2 g_{1}^{\text{in}}}{\text{d} \xi^2}  + \epsilon^2 \dfrac{\text{d}^2g_{2}^{\text{in}}}{\text{d} \xi^2} + \dots
\label{innersolutionderivatives}
    \end{cases}
\end{equation}
into Eq.(\ref{inner_equation_final}). Then, we apply the same procedure that we used for the outer solution: we group together all the terms linked with the same power of $\epsilon$ and we set them separately equal to zero. In the case of the first order term this means
\begin{multline}
    \dfrac{\text{d}^2 g_{1}^{\text{in}}}{\text{d} \xi^2} - 4 \dfrac{\langle \tilde{\textbf{v}}(\xi)\rangle}{G(\xi)} \dfrac{\text{d} g_{1}^{\text{in}}}{\text{d} \xi} + \biggl[ \biggl( \dfrac{2}{(\xi \epsilon + \tilde{r}_c)} + W(\xi) +\dfrac{G_r}{G} \biggr) \dfrac{\text{d} g_{0}^{\text{in}}}{\text{d} \xi} + \\ - 4 \dfrac{\langle \tilde{\nabla}_{\xi} \cdot \tilde{\textbf{v}}(\xi) \rangle}{G(\xi)} g_0^{\text{in}(\xi)} \biggr] = 0.
\label{first_order_inner_equation}
\end{multline}

The above equation can be analytically solved through the mathematical steps reported in Appendix D, ending up with the following expression:
\begin{multline}
    g_{1}^{\text{in}}(\xi) = C_3 + \int_0^{\xi} \biggl \{ C_2 - \int_0^{\xi} \biggl[ \biggl( \dfrac{2}{(\xi \epsilon + \tilde{r}_c)} + W(\xi) + \dfrac{G_r}{G}\biggr) \dfrac{\text{d} g_{0}^{\text{in}}}{\text{d} \xi}+ \\ - 4 \dfrac{\langle \tilde{\nabla}_{\xi} \cdot \tilde{\textbf{v}}(\xi) \rangle}{G(\xi)} g_0^{\text{in}}(\xi) \biggr] \exp \biggl(\int_0^{\xi} -4 \dfrac{\langle \tilde{\textbf{v}}(\xi)\rangle}{G(\xi)} d\xi \biggr) d\xi \biggr \} \times \\ \times \exp \biggl(\int_0^{\xi} 4 \dfrac{\langle \tilde{\textbf{v}}(\xi)\rangle}{G(\xi)} d\xi \biggr) d\xi.
\label{first_order_inner_solution}
\end{multline}

\subsection{Integration constants evaluation}
\begin{figure*}
	\centering
	\includegraphics[width = 0.6 \linewidth]{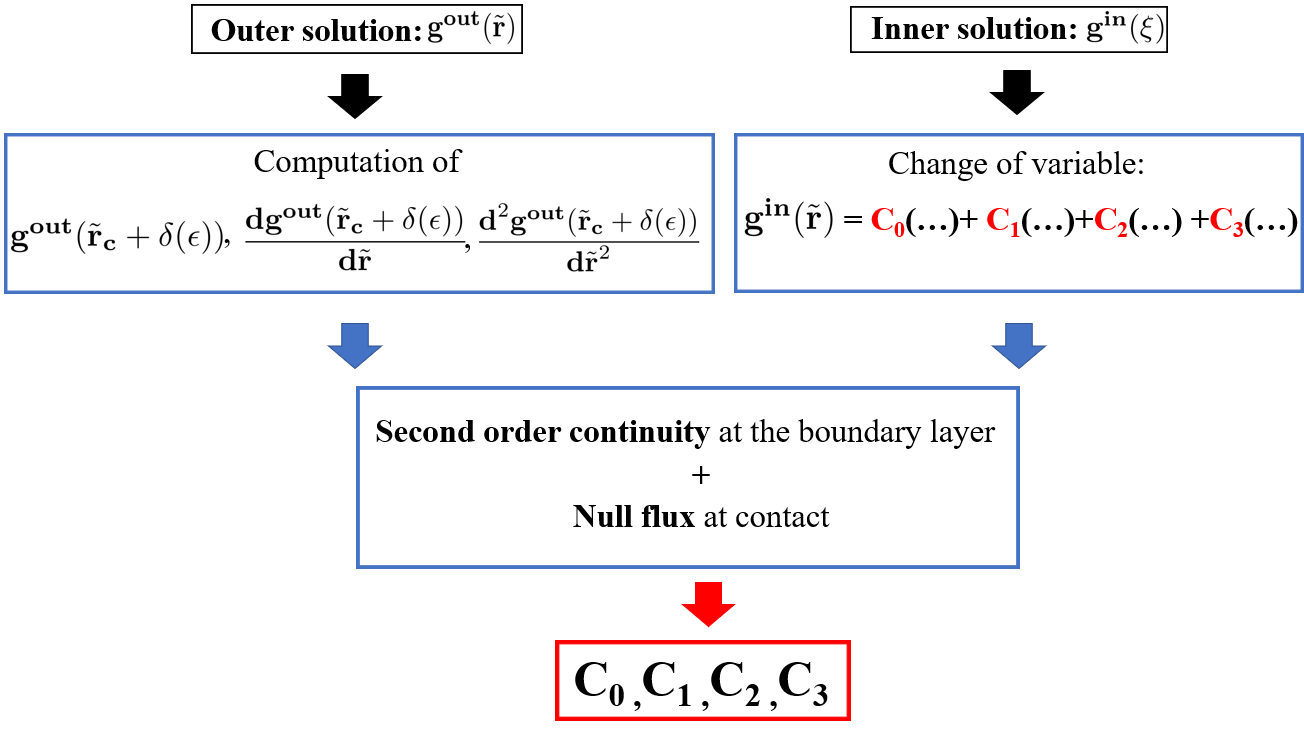}
	\caption{Block diagram with the fundamental steps for the evaluation of the integration constants within $g^{\text{in}}(\tilde{r})$}
	\label{fig:integration_constants_evaluation}
\end{figure*}

To summarize, we have evaluated two different series $g^{\text{in}}$ and $g^{\text{out}}$ which describe the behavior of the solution in two different adjacent sections of the integration domain.
The final step to obtain the analytical solution of Eq.(\ref{Smoluchowski_epsilon_final}) is the evaluation of the integration constants $C_0$, $C_1$, $C_2$ and $C_3$ present in the inner solution; the full procedure is summarized in Fig. \ref{fig:integration_constants_evaluation}.
Since we have four unknown parameters we need four equations to determine them: the first one will be the condition of zero flux at the reference particle surface, Eq.(\ref{BC_2}), while the other three can be obtained from the $\textit{patching}$ procedure \cite{BenderOrszag}.

The general principle is as follows. We start from two solutions which share a common border: if one of the two is known and the other has $N$ constants to be evaluated, it is necessary to apply a continuity of order $N-1$.\\
This principle is suitable for our case since we know the full behavior of the outer solution and we have three remaining conditions to be fixed in order to find the three remaining constants. Hence, we need to fix a second order continuity between $g^{\text{out}}$ and $g^{\text{in}}$ at their shared border, that is 
$\tilde{r} = \tilde{r}_c + \epsilon$.\\
After having obtained the complete structure of the inner solution, we need to group together all the coefficients related to each integration constant; for clarity we will adopt the following mathematical notation to describe the structure of the inner solution:
\begin{equation}
\begin{cases}
\text{Int}_0(\xi) = \exp{\int_0^{\xi} 4 \dfrac{\langle \tilde{\textbf{v}}(\epsilon = 0) \rangle}{G(\epsilon = 0)}} \text{d} \xi \\ 
\text{IntInt}_0(\xi) = \int_0^{\xi} \text{Int}_0(\xi) \text{d} \xi \\
\text{Int}_1(\xi) = \exp{\int_0^{\xi} 4 \dfrac{\langle \tilde{\textbf{v}}(\xi) \rangle}{G(\xi)}} \text{d} \xi \\ 
\text{T}(\xi) = 4 \dfrac{\langle  \tilde{\nabla}_{\xi} \cdot \tilde{\textbf{v}} \rangle}{G(\xi) \text{Int}_1(\xi)}\\
\text{Z}(\xi) =  4 \dfrac{\langle  \tilde{\nabla}_{\xi} \cdot \tilde{\textbf{v}} \rangle}{G(\xi)}\\
\text{Q}(\xi)  = \biggl[\text{T}(\xi) \text{IntInt}_0 - \biggl( \dfrac{2}{\xi \epsilon +\tilde{r}_c} + W(\xi) + \dfrac{G_r(\xi)}{G(\xi)} \biggr)\text{Int}_0 \biggr]\times\\
\quad \quad \quad \quad \times (\text{Int}_1)^{-1}\\
\text{R}(\xi) = \text{Z}(\xi)\text{IntInt}_0 - \biggl( \dfrac{2}{\xi \epsilon +\tilde{r}_c}+ W(\xi) + \dfrac{G_r(\xi)}{G(\xi)} \biggr) \text{Int}_0.
\end{cases}
\end{equation}

So, $g^{\text{in}}(\xi)$ can be written as
\begin{multline}
g^{\text{in}}(\xi) = C_0 \biggl \{  \text{IntInt}_0(\xi)+ \epsilon \biggl[ \int_0^{\xi} \biggl( \int_0^{\xi} \text{Q}(\xi) \text{d} \xi \biggr) \times \\ \times \text{Int}_1 \text{d} \xi\biggr] \biggr\} + C_1 \biggl \{ 1+ \epsilon \biggl[ \int_0^{\xi} \biggl( \int_0^{\xi} \text{T}(\xi) \text{d} \xi \biggr) \text{Int}_1 \text{d} \xi \biggr] \biggr \} + \\ + C_2 \delta(\epsilon) \int_0^{\xi} \text{Int}_1(\xi) \text{d} \xi+ C_3 \epsilon.
\end{multline}
To find the integration constants it is necessary to go back to the radial coordinate $\tilde{r}$:
\begin{multline}
	g^{\text{in}}(\tilde{r}) = C_0 \biggl \{  \text{IntInt}_0(\tilde{r})+ \dfrac{1}{\epsilon} \biggl[ \int_{\tilde{r}_c}^{\tilde{r}} \biggl( \int_{\tilde{r}_c}^{\tilde{r}} \text{Q}(\tilde{r}) \text{d} \tilde{r} \biggr) \times \\ \times \text{Int}_1 \text{d} \tilde{r} \biggr] \biggr\} + C_1 \biggl \{ 1+  \dfrac{1}{\epsilon} \biggl[ \int_{\tilde{r}_c}^{\tilde{r}} \biggl( \int_{\tilde{r}_c}^{\tilde{r}} \text{T}(\tilde{r}) \text{d}\tilde{r} \biggr) \text{Int}_1 \text{d} \tilde{r} \biggr] \biggr \} + \\ + C_2 \int_{\tilde{r}_c}^{\tilde{r}} \text{Int}_1(\tilde{r}) \text{d} \tilde{r} + C_3 \epsilon
\end{multline}
Next, we need to write the first and the second order derivative of $g^{\text{in}}(\tilde{r})$:
\begin{multline}
\dfrac{\text{d} g^{\text{in}}(\tilde{r})}{\text{d} \tilde{r}} = C_0 \biggl[ \text{Int}_0(\tilde{r})  + \dfrac{1}{\epsilon} \biggl( \int_{\tilde{r}_c}^{\tilde{r}} \text{Q}(\tilde{r}) \text{d} \tilde{r} \biggr)  \text{Int}_1(\tilde{r}) \biggr] + \\ +  C_1 \biggl[  \dfrac{1}{\epsilon} \biggl( \int_{\tilde{r}_c}^{\tilde{r}} \text{T}(\tilde{r} ) \text{d} \tilde{r} \biggr) \text{Int}_1(\tilde{r}) \biggr] + C_2 \text{Int}_1(\tilde{r})
\end{multline}
\begin{multline}
\dfrac{\text{d}^2 g^{\text{in}}(\tilde{r})}{\text{d} \tilde{r}^{2}} = C_0 \biggl[ \dfrac{ \text{d} \text{Int}_0(\tilde{r})}{\text{d} \tilde{r}}  + \dfrac{1}{\epsilon} \biggl( \int_{\tilde{r}_c}^{\tilde{r}} \text{Q}(\tilde{r}) \text{d} \tilde{r} \biggr)  \dfrac{\text{d} \text{Int}_1(\tilde{r}) }{\text{d} \tilde{r}}  + \\ + R(\tilde{r}) \biggr] +  C_1 \biggl[ \dfrac{1}{\epsilon} \biggl( \int_{\tilde{r}_c}^{\tilde{r}} \text{T}(\tilde{r}) \text{d} \tilde{r} \biggr) \dfrac{\text{d} \text{Int}_1(\tilde{r})}{\text{d} \tilde{r}} + Z(\tilde{r}) \biggr] + \\+ C_2 \dfrac{\text{d} \text{Int}_1(\tilde{r})}{\text{d} \tilde{r}}
\end{multline}
It is possible to evaluate all the integration constants through the resolution of the following linear system which is dictated by the patching procedure,
\begin{equation}
\begin{cases}
	g^{\text{out}}(\tilde{r} = \tilde{r}_c + \epsilon) = g^{\text{in}}(\tilde{r} = \tilde{r}_c + \epsilon) \\
	\dfrac{\text{d} g^{\text{out}}(\tilde{r} = \tilde{r}_c + \epsilon)}{\text{d} \tilde{r}} = \dfrac{\text{d} g^{\text{in}}(\tilde{r} = \tilde{r}_c + \epsilon)}{\text{d} \tilde{r}} \\
\dfrac{\text{d}^2 g^{\text{out}}(\tilde{r} = \tilde{r}_c + \epsilon)}{\text{d} \tilde{r}^2} = \dfrac{\text{d}^2 g^{\text{in}}(\tilde{r} = \tilde{r}_c + \epsilon)}{\text{d} \tilde{r}^2}
\end{cases}
\label{Integration_Constants_Evaluation}
\end{equation}
together with the application of Eq.(\ref{BC_2}) recalled briefly here:
\begin{multline}
    0 = G(\tilde{r}_c)  \dfrac{\text{d}g^{\text{in}}}{\text{d}\tilde{r}}(\tilde{r}_c)  + \\ + \biggl( G(\tilde{r}_c) \dfrac{\text{d} \tilde{U}}{\text{d}\tilde{r}}(\tilde{r}_c) - 4 \text{Pe} \langle \tilde{\textbf{v}} \rangle \biggr) g^{\text{in}}(\tilde{r}_c).
\end{multline}
The  result of Eq.(\ref{Integration_Constants_Evaluation}) and Eq.(\ref{BC_2}) is the evaluation of the four integration constants $C_0$, $C_1$, $C_2$ and $C_3$ as functions of the P\'eclet number which will provide the final form of $g^{\text{in}}$.\\

In the following section we will present the results related to the spherically-averaged rdf which will be given by the patching of the inner solution inside the boundary layer $\delta(\epsilon)$ with the outer solution outside the boundary layer:
\begin{equation}
\begin{cases}
	g(\tilde{r}) = g_0^{\text{in}} + \epsilon g_1^{\text{in}} \quad \tilde{r} < \text{r}_c + \epsilon\\
	g(\tilde{r}) = g_0^{\text{out}} + \epsilon g_1^{\text{out}} \quad \tilde{r} \ge \text{r}_c + \epsilon.
\label{spherical_averaged_pair_correlation_function}

\end{cases}
\end{equation}

\section{Results}
\begin{figure*}
	\centering
	\subfloat{\includegraphics[width = 0.40 \linewidth]{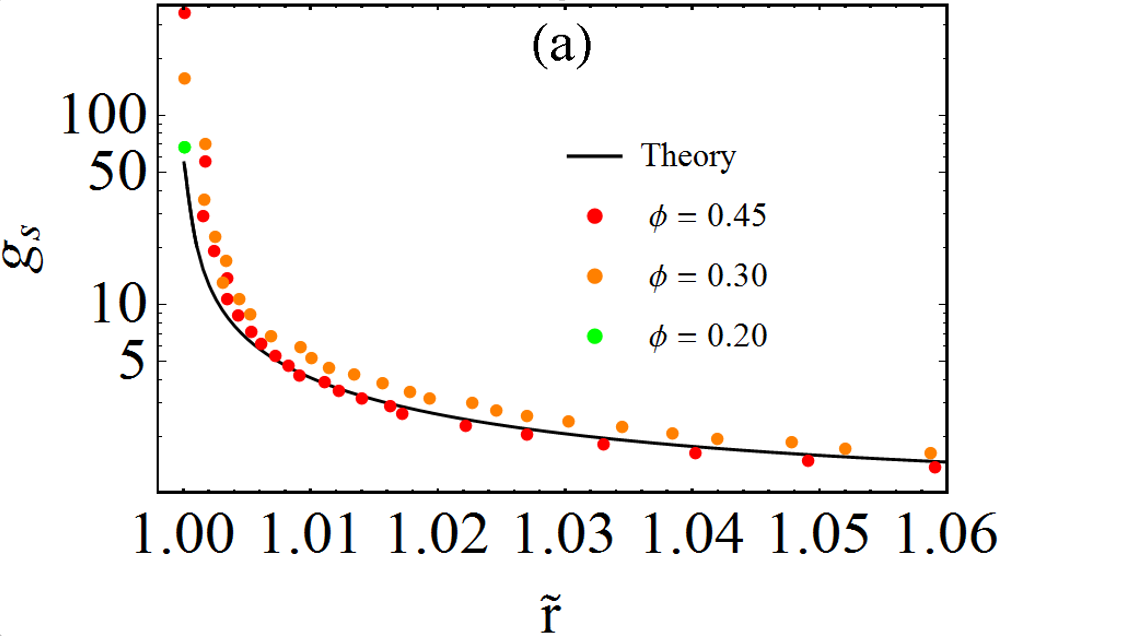}} \quad
	\subfloat{\includegraphics[width = 0.38 \linewidth]{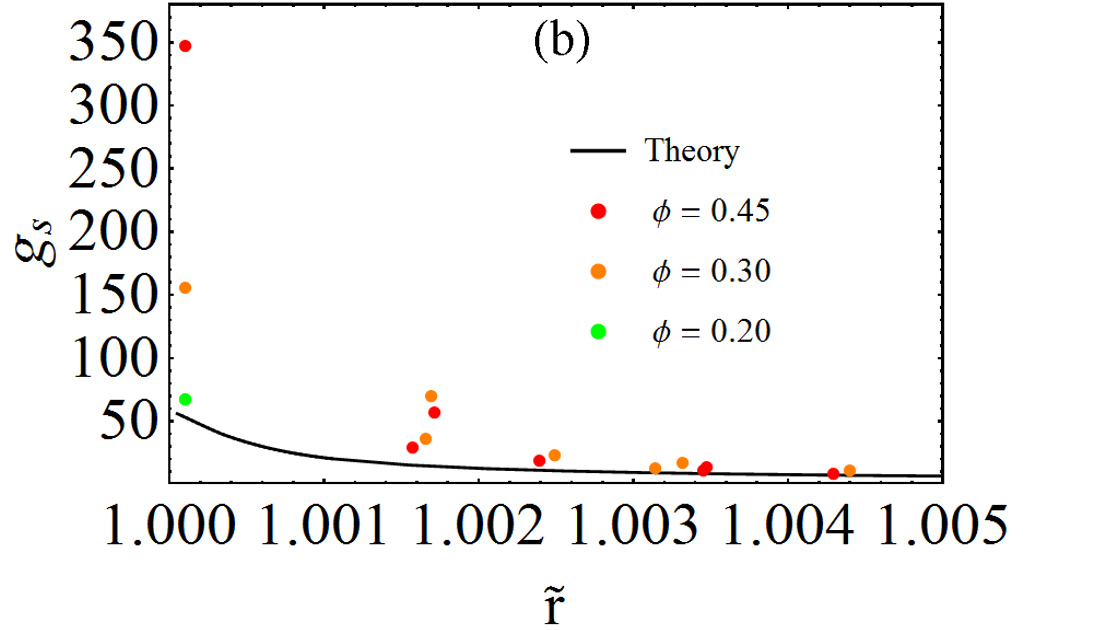}}  \quad
	\caption{Comparison of  $\text{g}_s(\tilde{r})$ evaluated from Eq. (\ref{spherical_averaged_pair_correlation_function})  with simulations (symbols) of Hard-Sphere 	    suspensions at Pe = 1000 \cite{Morris} at different volume fractions represented in logarithmic (a) and linear scale (b).}
	\label{fig:convalida}
\end{figure*}
Our main interest is the analysis of the rdf inside the compression quadrants, which contains the more interesting physics, where $\phi \in [\pi/2,\pi] $ and $ \phi\ \in [3\pi/2,2\pi]$. On the other hand, the only available data to verify this analytical approach are given by simulations of hard-spheres where the rdf is evaluated for $\phi \in [0,2\pi]$.
For this reason and for validation purposes, we include calculations for $\text{Pe} = 1000$ related to the extensional quadrants, where $V_r >0$, into the calculations and then average over the solid angle to obtain the rdf averaged over all sectors. From this point onward we will refer to the rdf averaged over all $\phi$ as $g_s(\tilde{r})$, to be distinguished from the rdf averaged only over the compressional sectors, which will be denoted as $g_c(\tilde{r})$.
\subsection{Hard Spheres}
This case corresponds to setting $\tilde{U}(\tilde{r}) = 0$ in the above derivation.

\subsubsection{\textbf{Comparison with simulation data for Hard Spheres}}
In Fig.\ref{fig:convalida} we compare the behavior of the $\text{g}_s(\tilde{r}$) with the results obtained through Stokesian dynamics simulations \cite{Morris} for strongly sheared suspensions at different volume fractions; we also plot an additional point obtained from numerical simulations from the same contest. 

It is seen that, at  distances from the surface of the reference particle up to $\tilde{r} = 1.0015$, the match between the theory and all the sets of simulation data is good. Afterwards, both the theory and the simulations shows a peak. We observe first that the peak height provided by the theory is always significantly below the simulation data for $\phi = 0.45$ and $\phi = 0.30$, which is reasonable because our theory applies to the dilute or at most semi-dilute regime. As a confirmation of this, the height of the peak provided by the theory is instead very close to the peak value of simulations data at $\phi = 0.20$.

As a conclusion, Fig.3 confirms the validity of the theory, where no fitting or adjustable parameters are used. This  prediction for the Hard-Sphere case represents also an improvement with respect to previous theories \cite{MorrisBrady} where the prediction of the rdf near contact with the reference particle was somewhat overestimated.

\subsubsection{\textbf{Effect of the P\'eclet number}}
\begin{figure}
	\centering
	\subfloat{\includegraphics[width = 0.46 \linewidth]{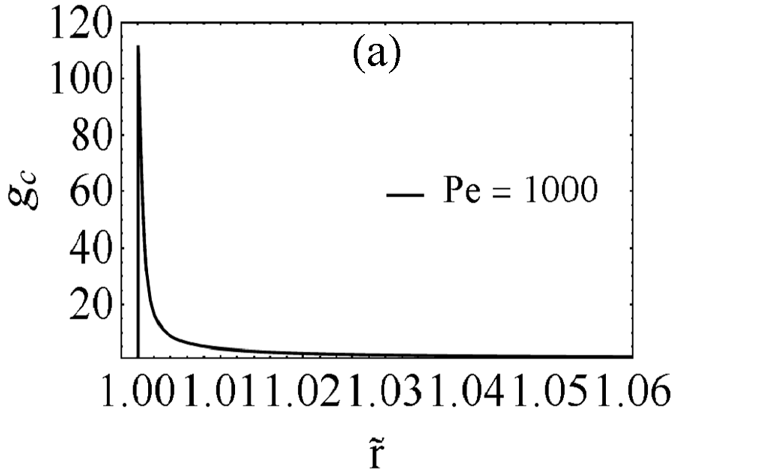}} \quad
	\subfloat{\includegraphics[width = 0.45 \linewidth]{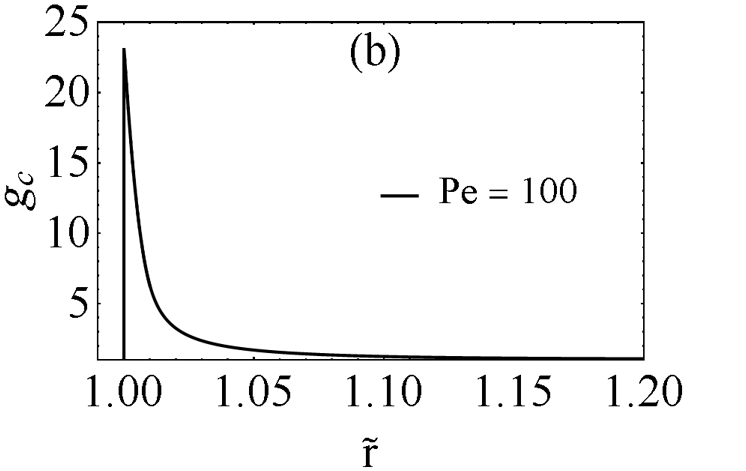}}  \quad
	\subfloat{\includegraphics[width = 0.45 \linewidth]{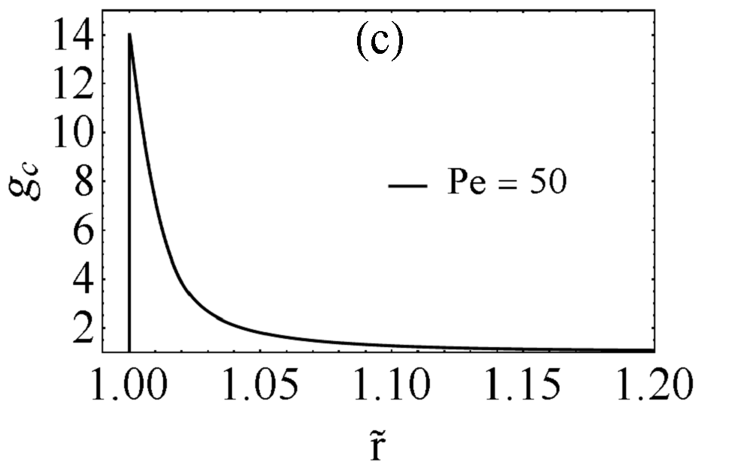}}  \quad
	\subfloat{\includegraphics[width = 0.45 \linewidth]{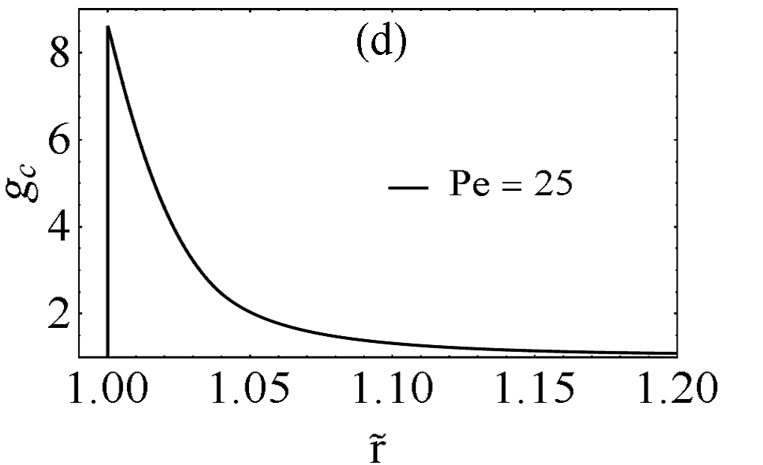}}  \quad
	\subfloat{\includegraphics[width =  0.50 \linewidth]{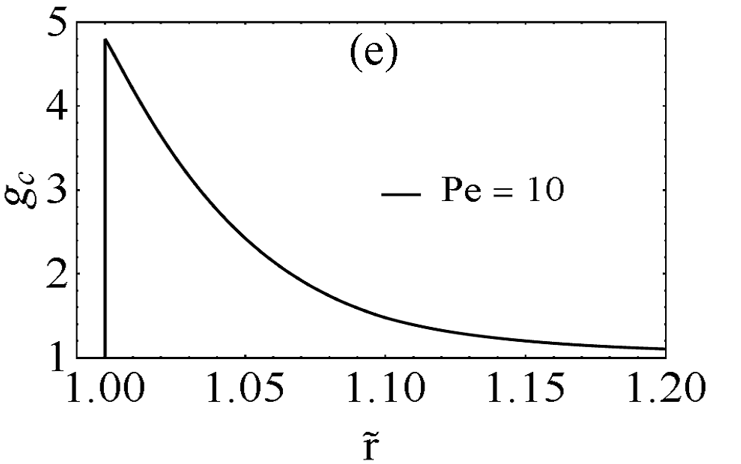}}   \quad
	\caption{Effect of the P\'eclet number on the trend of  $\text{g}_c(\tilde{r})$ for Hard-Sphere models.}
	\label{fig:pair_correlation_compressing}
\end{figure}

We are interested in determining how the rdf evolves as a function of the P\'eclet number. \\
In Fig. \ref{fig:pair_correlation_compressing} the behaviors evaluated for the Hard-Sphere model with different values of $\text{Pe}$ are plotted. We can immediately  notice that the peak of the $g_{\text{c}}(\tilde{r})$ decreases with the P\'eclet number: this statement is physically meaningful because, as the compressing effect provided by the advective term is getting weaker, the probability to find the second particle close to the reference one decreases. Moreover, upon decreasing the P\'eclet number, the region where the balance between the advective and the Brownian contribution is not negligible increases: this is also reflected by the increasing width of the boundary layer $\delta$ which is inversely proportional to the P\'eclet number.
\subsection{Radial Distribution Function of the Lennard-Jones fluid under shear flow}
Finally, we consider the influence of a non-trivial interaction potential on the rdf of the sheared fluid. 
In particular, we will focus on the the 12-6 Lennard-Jones (LJ) interaction potential, which forms a paradigm for many liquids:
\begin{equation}
    \tilde{U}(\tilde{r}) = \tilde{U}_{\text{min}} \biggl( \tilde{r}^{-12} -\tilde{r}^{-6} \biggr) = 4 \lambda \biggl( \tilde{r}^{-12} -\tilde{r}^{-6} \biggr),
    \label{L_J_12_6} 
\end{equation}
where $\lambda$ is the minimum of the interaction potential between the particles.

\begin{figure}
	\centering
	\includegraphics[width = 0.9 \linewidth]{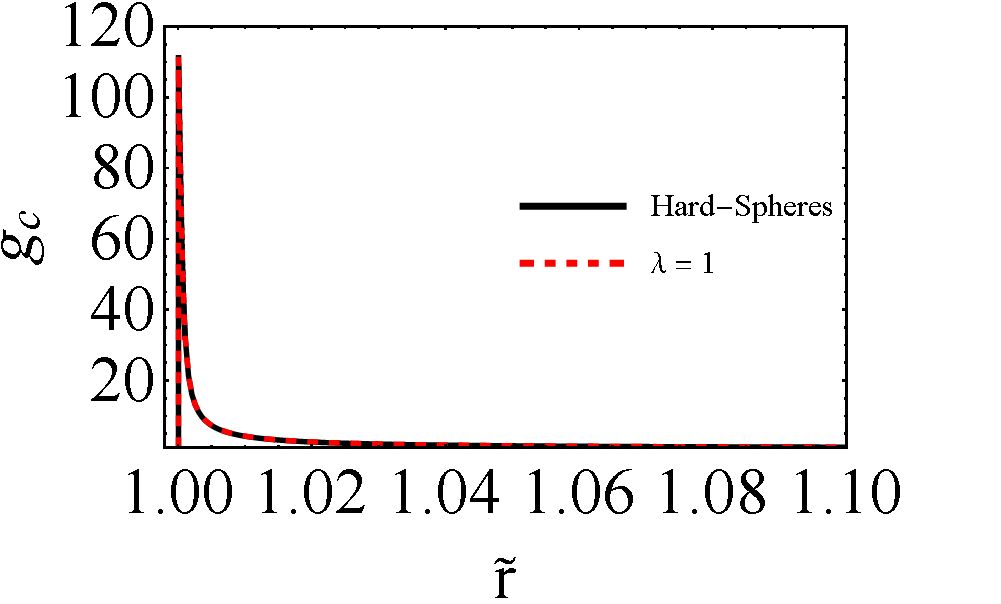}
	\caption{Effect of the interaction potential on $g_c(\tilde{r})$ at high values of the P\'eclet number (Pe = 1000).}
	\label{fig:High_Pe_Interaction}
\end{figure}

It is expected that, for large values of the P\'eclet number, the effect of the LJ potential is totally negligible since the shear flow contribution is dominant. This feature is clearly demonstrated in Fig.\ref{fig:High_Pe_Interaction} where the $g_c(\tilde{r})$ for the calculation including the LJ potential is exactly the same as for the hard sphere calculation.

\begin{figure}
	\centering
	\includegraphics[width = 0.9 \linewidth]{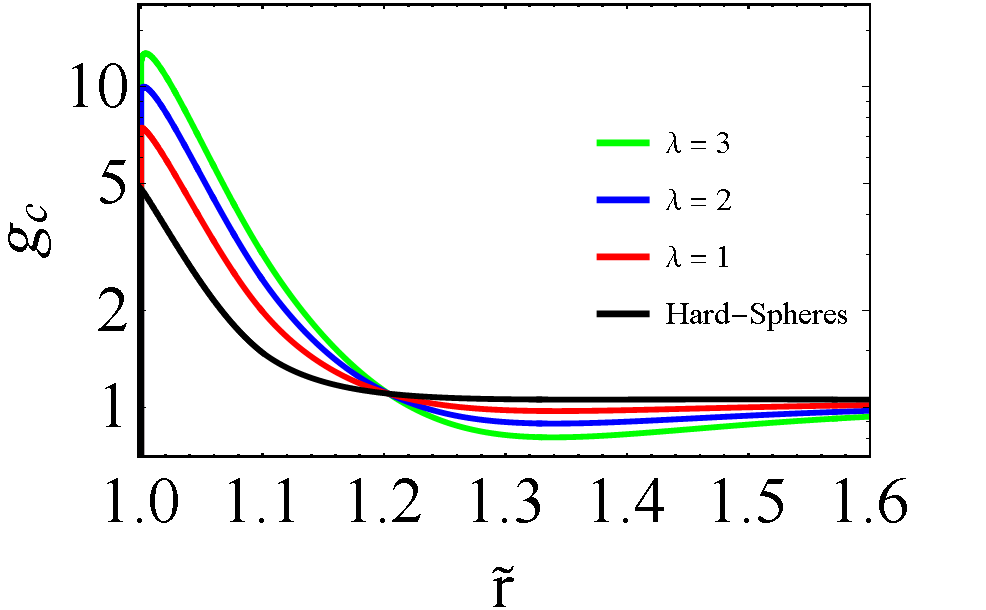}
	\caption{ $g_c(\tilde{r})$ at low values of the P\'eclet number (Pe = 10) for an hard-sphere suspension (lowest curve) and different values of $\lambda$ (at an higher curve corresponds a  bigger value of $\lambda$).}
	\label{fig:Low_Pe_Interaction}
\end{figure}

On the other hand, if the P\'eclet number is sufficiently small, as the one chosen for Fig \ref{fig:Low_Pe_Interaction}, we can clearly see the effect of the interaction potential at different values of $\lambda$. To this aim, we can divide the radial domain into three different regions.\\

In the radial region furthest away from the reference particle, we see a significant undershoot and a minimum in $g_c(\tilde{r})$, which becomes deeper upon increasing the attraction energy of the LJ potential. 
Interestingly, this undershoot is absent in the hard sphere case, and is an original prediction of the theory developed here for the LJ fluid. From a physical point of view, this minimum represents a depletion region which is necessary in order to balance (from a continuity point of view) the strong accumulation of particles at shorter distance caused by the synergy between the attractive force of the LJ and the action of the shear flow, both of which are pushing the particles close to the reference one. Indeed, at closer distance we find a broad and pronounced peak which reflects the increased probability of finding particles in that region, due to the action of both the attractive interaction and the shear flow. 

In other words, there is a region corresponding to the peak where the particles are pushed towards the reference one so strongly that there must be a nearby region at further distance where particles are depleted.
Interestingly, we also note that with the LJ potential the accumulation peak becomes smooth and non-singular, in contrast with the peak of the hard sphere system, which, instead, features a singularity at the point of maximum of the peak. This feature is to be attributed to the softness of the short-range repulsive part of the interaction of the LJ potential.

It is important to highlight that the depth of the undershoot, the height of the peak and the slope of $g_c(\tilde{r})$ before and after the maximum, all increase with the attraction parameter $\lambda$: a deeper well of potential corresponds to stronger attractive long-range interactions and also, at the same time, to stronger short-range repulsive contributions.\\

It is also important to notice the following feature: usually, if a soft sphere potential is used to describe the inter-particle interactions, then the radial distribution function goes to zero at values $\tilde{r} \le 1$, and for the LJ potential this happens at $\tilde{r}=0$. In our case the asymptotic behavior as $\tilde{r}\to 1$ is different, because of the effect of hydrodynamic interactions, including lubrication forces which are described by $G(\tilde{r})$. This function has to go to zero at the solid-liquid interface, i.e. right at the particle surface, at $\tilde{r}=1$ because of the divergence of the hydrodynamic lubrication resistance at the particle surface.
Had we considered the particles in vacuo or in a gaseous phase, then the rdf would tend to zero at 
$\tilde{r}_c \to 0$.

This can be seen more quantitatively by considering the second boundary condition Eq.\ref{BC_2}:
\begin{multline}
 0 = G(\tilde{r}_c)  \dfrac{\text{d}g^{\text{in}}}{\text{d}\tilde{r}}(\tilde{r}_c)  + \\ + \biggl( G(\tilde{r}_c) \dfrac{\text{d} \tilde{U}}{\text{d}\tilde{r}}(\tilde{r}_c) - 4 \text{Pe} \langle \tilde{\textbf{v}} \rangle \biggr) g^{\text{in}}(\tilde{r}_c).
\end{multline}
At the particle surface $\tilde{r}_c = 1$ then $G(\tilde{r}_c)$ is null because the hydrodynamic resistance diverges there; if we consider that the derivatives of both the rdf and the interaction potential cannot be zero at $\tilde{r}_c = 1$ we end up with
\begin{equation}
	- 4 \text{Pe} \langle \tilde{\textbf{v}} \rangle g^{\text{in}}(\tilde{r}_c) = 0
\end{equation}
Since we know that both the P\'eclet number and average radial velocity are not null, then the rdf at 
$\tilde{r}_c = 1$ must be zero.

\section{Conclusion}
In this work we have proposed a method to analytically solve the two body Smoluchowski equation with shear flow in a spherical reference frame by means of an intermediate asymptotics method.
The result of this work is the evaluation of the radial distribution function (rdf), which describes the microstructure of colloidal suspensions under a simple shear flow. The reliability of the method across a wide range of conditions of P\'eclet numbers and in analyzing the influence of an attractive interaction potential on the behaviour of the rdf has been demonstrated. 
The theory predicts an important feature of the rdf of attractive fluids in compressive quadrants: the presence of a pronounced depletion effect resulting in a minimum or undershoot in the rdf at separations right after the accumulation peak. This effect is already visible at modest values of the attraction energy, and becomes more and more important upon increasing the attraction. This new effect may have important consequences on the rheology (e.g. viscosity, non-Newtonian behaviors etc) of colloidal suspensions, which will be explored in future studies.

The method presented here is fairly general and can easily be extended in future work to systems of great interest such as one component plasmas \cite{Hansen,Larsen}, the rdf of denser and ordered systems ~\cite{Yurchenko}, droplet clustering in atmospheric flows~\cite{Glienke}, and nucleation and crystallization under shear flows~\cite{Mura}.

\section{Acknowledgements}
M. Morbidelli is gratefully acknowledged for many inspiring discussions and for providing motivation to study this problem. L.B. gratefully acknowledges financial support from Synthomer UK Ltd.

\appendix
\section{Angular averaging}
In this section we describe the procedure where we describe the angular averaging procedure with which we  evaluate $\langle \tilde{\textbf{v}} \rangle$ and $\langle  \tilde{\nabla} \cdot \tilde{\textbf{v}} \rangle$.
First we introduce the spherical components of the dimensionless velocity $\tilde{\textbf{v}}$ in a simple shear flow whose equivalent Cartesian coordinates are ($\dot{\gamma} y, 0 ,0$) \cite{shear_flow_spherical_coordinates}:
\begin{equation}
    \begin{cases}
        \tilde{v}_r = \tilde{r} (1-A(\tilde{r}))\sin^2{\theta} \sin\phi \cos\phi \\
        \tilde{v}_{\theta} = \tilde{r} (1- B(\tilde{r})) \sin{\theta} \cos{\theta}\sin\phi \cos\phi,\\
        \tilde{v}_{\phi} = \tilde{r} \sin\theta\biggl( \cos^2\phi - \dfrac{B(\tilde{r})}{2} \cos(2 \phi) \biggr)
    \end{cases}
    \label{dimensionless_velocity_axisymmetric}
\end{equation}
where $A(\tilde{r})$ and $B(\tilde{r})$ are functions representing the effect of the hydrodynamic disturbance along the radial and angular coordinate, respectively. Their values can be taken from the literature \cite{Batchelor} and, in order to use them in the present analytical calculations, they are fitted through the following algebraic expressions \cite{Melis}:
\begin{equation}
\begin{cases}
A(\tilde{r}) = \dfrac{113.2568894}{(2 \tilde{r})^5} +\dfrac{307.8264828}{(2 \tilde{r})^6} +\\
 - \dfrac{2607.54064288}{(2 \tilde{r})^7} + \dfrac{3333.72020041}{(2 \tilde{r})^8} \\\\
 
B(\tilde{r}) = \dfrac{0.96337157}{(2 \tilde{r} - 1.90461683)^{1.99517070}} +\\
- \dfrac{0.93850774}{(2 \tilde{r} - 1.90378420)^{2.01254004}}.
\label{hydrodynamic_functions_fitting}
\end{cases}
\end{equation}
Our goal is to evaluate the average radial velocity in the area where the particles are approaching each other, which means the ensemble of angular coordinates $\tilde{v}_r < 0 $.\\

It is found that the above mentioned condition is satisfied, for $\tilde{r} >0$, $\forall \theta \in [0,\pi]$, $\phi \in [\pi/2,\pi]$ and $\phi \in [3\pi/2, 2\pi]$. Now we apply the angular average obtaining:
\begin{multline}
     \langle \tilde{\textbf{v}} \rangle =  \tilde{r} (1-A(\tilde{r}))  \dfrac{1}{4 \pi}  \biggl[\int_0^{\pi} \sin^2(\theta) \sin(\theta) d\theta \times \\ \times \biggl( \int_{\pi/2}^{\pi} \sin(\phi) \cos(\phi) d \phi + \int_{3\pi/2}^{2 \pi} \sin(\phi) \cos(\phi) d \phi \biggr) \biggr].
    \label{angular_average_relative_velocity_compression}
\end{multline}
Through this procedure we can obtain
\begin{equation}
    \alpha_c = -\dfrac{1}{3 \pi}.
\end{equation}
To find the upstream region we need to impose $\tilde{v}_r>0$, which is given by  $\forall \theta \in [0,\pi]$, $\phi \in [0,\pi/2]$ and $\phi \in [\pi, 3\pi/2]$. Applying the same procedure seen before for $\alpha_c$ we obtain:
\begin{multline}
     \langle \tilde{\textbf{v}} \rangle =  \tilde{r} (1-A(\tilde{r}))  \dfrac{1}{4 \pi}  \biggl[\int_0^{\pi} \sin^2(\theta) \sin(\theta) d\theta \times \\ \times \biggl( \int_{0}^{\pi/2} \sin(\phi) \cos(\phi) d \phi + \int_{\pi}^{ 3\pi/2} \sin(\phi) \cos(\phi) d \phi \biggr) \biggr],
    \label{angular_average_relative_velocity_extension}
\end{multline}
and, as a consequence
\begin{equation}
    \alpha_e = \dfrac{1}{3 \pi}.
\end{equation}
From this point onward we will consider the compressional case only; the extensional one can be derived in a straightforward manner by replacing $\alpha_c$ with $\alpha_e$.

Next we consider the divergence of the actual flow field, which can be written in polar coordinates as
\begin{multline}
    \tilde{\nabla} \cdot \tilde{\textbf{v}} = \\ =\dfrac{1}{\tilde{r}^2} \dfrac{\partial}{\partial \tilde{r}} \biggl( \tilde{r}^2 \tilde{v}_r \biggr) + \dfrac{1}{\tilde{r} \sin({\theta})} \dfrac{\partial}{\partial \theta} \biggl( \sin({\theta}) v_{\theta} \biggr) + \dfrac{1}{\tilde{r} \sin{\theta}} \dfrac{\partial}{\partial \phi}\biggl( v_{\phi} \biggr).
\end{multline}
Adopting the correlations in Eq.(\ref{dimensionless_velocity_axisymmetric}), we can evaluate the divergence as
\begin{equation}
    \tilde{\nabla} \cdot \tilde{\textbf{v}} = \biggl[ 3 (B(\tilde{r})-A(\tilde{r}))-\tilde{r} \dfrac{\text{d}A}{\text{d}\tilde{r}} \biggr] \sin^2{\theta} \sin{\phi} \cos{\phi}.
\end{equation}
Finally, we apply the integral average previously seen for the radial velocity and we obtain:
\begin{equation}
     \langle  \tilde{\nabla} \cdot \tilde{\textbf{v}} \rangle = \alpha_c \biggl[ 3 (B(\tilde{r})-A(\tilde{r}))-\tilde{r} \dfrac{\text{d}A}{\text{d}\tilde{r}} \biggr].
     \label{angular_averaged_divergence}
\end{equation}
\section{Outer solution calculations}
To evaluate $g_0^{\text{out}}$ we can express Eq.(\ref{leading_order_solution}) as
\begin{equation}
   \dfrac{\text{d} g_0^{\text{out}}(\tilde{r})}{\text{d} r} = - g_0^{\text{out}}(\tilde{r}) \dfrac{\langle \tilde{\nabla} \cdot \textbf{v} \rangle }{ \langle \tilde{\textbf{v}} \rangle }.\\
   \label{leading_order_solution_I}
\end{equation}
Replacing Eq.(\ref{angular_average_relative_velocity_compression}) and Eq.(\ref{angular_averaged_divergence}) we obtain
\begin{equation}
    \dfrac{\text{d} g_0^{\text{out}}(\tilde{r})}{g_0^{\text{out}}} = - \biggl[\dfrac{3 (B(\tilde{r})-A(\tilde{r}))}{\tilde{r} (1-A(\tilde{r}))} - \dfrac{\text{d}A/\text{d}\tilde{r}}{1-A(\tilde{r})}\biggr] \text{d} r.\\
\label{leading_order_solution_II}
\end{equation}
Integrating Eq.(\ref{leading_order_solution_II}) within the range [$\tilde{r}$,$\infty$] we obtain
\begin{multline}
    \ln{(g_0^{\text{out}}(\tilde{r} \to \infty))} - \ln{(g_0^{\text{out}}(\tilde{r}))} = \\ = \int_{\tilde{r}}^{\infty}   \biggl[\dfrac{3 (B(\tilde{r})-A(\tilde{r}))}{\tilde{r} (1-A(\tilde{r}))} + \dfrac{\text{d}A/\text{d}\tilde{r}}{1-A(\tilde{r})}\biggr] \text{d} r,
\end{multline}
ending up with the final form
\begin{equation}
    g_0^{\text{out}}(\tilde{r}) = \dfrac{1}{1-A} \exp \biggl[ \int_{\tilde{r}}^{\infty}  \dfrac{3 (B-A)}{\tilde{r} (1-A)} \text{d} \tilde{r} \biggr].
\end{equation}
Next, we evaluate $g_1^{\text{out}}(\tilde{r})$ starting from Eq.(\ref{firstorderouterequation}):
\begin{multline}
    \biggl[  G \biggl( \dfrac{\text{d}^2  g_{0}^{\text{out}}}{\text{d} \tilde{r}^2}  + \dfrac{2}{\tilde{r}} \dfrac{\text{d}  g_{0}^{\text{out}}}{\text{d}\tilde{r}}\biggr) + \dfrac{\text{d} G}{\text{d}\tilde{r}} \dfrac{\text{d}  g_{0}^{\text{out}}}{\text{d}\tilde{r}}+ g_{0}^{\text{out}} \dfrac{\text{d} \tilde{U}}{\text{d} \tilde{r}} \dfrac{\text{d} G}{\text{d} \tilde{r}}+ \\ + G \dfrac{\text{d} \tilde{U}}{\text{d} \tilde{r}} \dfrac{\text{d}  g_{0}^{\text{out}}}{\text{d} \tilde{r}} + G \biggl( \dfrac{2}{\tilde{r}} \dfrac{\text{d}\tilde{U}}{\text{d} \tilde{r}} + \dfrac{\text{d}^2 \tilde{U}}{\text{d} \tilde{r}^2} \biggr) g_{0}^{\text{out}} \biggr] + \\ -  4 \biggl( \langle \tilde{\textbf{v}} \rangle \dfrac{\text{d} g_1^{\text{out}}}{\text{d} \tilde{r}} + g_{1}^{\text{out}} \langle \tilde{\nabla} \cdot \textbf{v} \rangle \biggr) = 0.
    \label{firstorderouterequation_I}
\end{multline}
It is possible to explicitly write the first and the second order derivative of $g_0^{\text{out}}$ as
\begin{equation}
\begin{cases}
    \dfrac{\text{d} g_0^{\text{out}}}{\text{d} \tilde{r}} = - \biggl[ \dfrac{3 (B(\tilde{r})-A(\tilde{r}))}{\tilde{r} (1-A(\tilde{r}))} - \dfrac{\text{d}A/\text{d}\tilde{r}}{1-A(\tilde{r})} \biggr] g_0^{\text{out}} = Y(\tilde{r}) g_0^{\text{out}} \\ \\ 
    \dfrac{\text{d}^2 g_0^{\text{out}}}{\text{d} \tilde{r}^2} = \biggl( Y(r)^2 + \dfrac{\text{d} Y(\tilde{r})}{\text{d} \tilde{r}} \biggr) g_0^{\text{out}}
\end{cases}
\end{equation}
which defines the coefficient $Y(\tilde{r})$ in the main text.

Replacing the equations in Eq.(\ref{firstorderouterequation_I}) we obtain
\begin{multline}
\dfrac{g_{0}^{\text{out}}(\tilde{r})}{4 \langle \tilde{\textbf{v}} \rangle} \biggl\{G(\tilde{r})\biggl(Y^2 + \dfrac{\text{d} Y}{\text{d} \tilde{r}} \biggr) + \biggl[ G(\tilde{r}) \biggl( \dfrac{2}{\tilde{r}}  + \dfrac{\text{d} \tilde{U}}{\text{d} \tilde{r}}\biggr) + \dfrac{\text{d}G}{\text{d}\tilde{r}} \biggr] Y(\tilde{r}) + \\ +G(\tilde{r})\biggl( \dfrac{\text{d}^2 \tilde{U}}{\text{d} \tilde{r}^2} + \dfrac{2}{\tilde{r}} \dfrac{\text{d}\tilde{U}}{\text{d} \tilde{r}} \biggr) + \dfrac{\text{d} G}{\text{d} \tilde{r}} \dfrac{\text{d}\tilde{U}}{\text{d}\tilde{r}} \biggr\} = \\ = \dfrac{\text{d}g_1^{\text{out}}}{\text{d} \tilde{r}} +  \biggl[ \dfrac{3 (B-A)}{\tilde{r}(1-A)}-\tilde{r} \dfrac{\text{d}A/\text{d}\tilde{r}}{(1-A)} \biggr].
\label{firstorderouterequation_II}
\end{multline}

Equation (\ref{firstorderouterequation_II}) can be solved analytically multiplying both sides by an integrating factor:
\begin{multline}
    \mu(\tilde{r}) = (1-A) \exp{ \biggl[ \int_{\tilde{r}}^{\infty} - \biggl( \dfrac{3 (B-A)}{\tilde{r} (1-A)} \biggr)\text{d} \tilde{r} \biggr] } = \\ = \dfrac{1}{g_0^{\text{out}}},
\end{multline}
which simplifies it as:
\begin{multline}
    \dfrac{\text{d}}{\text{d} \tilde{r}} \biggl( \mu(\tilde{r}) g_1^{\text{out}} \biggr) = \\= \dfrac{1}{4 \langle \tilde{\textbf{v}} \rangle} \biggl\{G(\tilde{r})\biggl(Y^2 + \dfrac{\text{d} Y}{\text{d} \tilde{r}} \biggr) + \biggl[ G(\tilde{r}) \biggl( \dfrac{2}{\tilde{r}}  + \dfrac{\text{d} \tilde{U}}{\text{d} \tilde{r}}\biggr) + \dfrac{\text{d}G}{\text{d}\tilde{r}} \biggr] Y(\tilde{r}) + \\ +G(\tilde{r})\biggl( \dfrac{\text{d}^2 \tilde{U}}{\text{d} \tilde{r}^2} + \dfrac{2}{\tilde{r}} \dfrac{\text{d}\tilde{U}}{\text{d} \tilde{r}} \biggr) + \dfrac{\text{d} G}{\text{d} \tilde{r}} \dfrac{\text{d}\tilde{U}}{\text{d}\tilde{r}} \biggr\}.
\end{multline}
Finally it is possible to integrate the previous equation within the interval [$\tilde{r}$,$\infty$] obtaining the final form of $g_1^{\text{out}}$:
\begin{multline}
g_1^{\text{out}}(\tilde{r}) = - g_0^{\text{out}} \int_{\tilde{r}}^{\infty}  \dfrac{1}{4 \langle \tilde{\textbf{v}} \rangle} \biggl\{G(\tilde{r})\biggl(Y^2 + \dfrac{\text{d} Y}{\text{d} \tilde{r}} \biggr) +\\+ \biggl[ G(\tilde{r}) \biggl( \dfrac{2}{\tilde{r}}  + \dfrac{\text{d} \tilde{U}}{\text{d} \tilde{r}}\biggr) + \dfrac{\text{d}G}{\text{d}\tilde{r}} \biggr] Y(\tilde{r}) +G(\tilde{r})\biggl( \dfrac{\text{d}^2 \tilde{U}}{\text{d} \tilde{r}^2} + \dfrac{2}{\tilde{r}} \dfrac{\text{d}\tilde{U}}{\text{d} \tilde{r}} \biggr) + \\ + \dfrac{\text{d} G}{\text{d} \tilde{r}} \dfrac{\text{d}\tilde{U}}{\text{d}\tilde{r}} \biggr\} \text{d} \tilde{r}.
\end{multline}
\section{Gauge functions}
In this appendix we will evaluate the asymptotic behavior of each function in Eq.(\ref{Smoluchowski_epsilon_inner}) in the limit $\epsilon \to 0$, to determine the gauge functions associated with every term in the equation.

At first, we will focus on the angular-averaged quantities:
\begin{equation}
\begin{cases}
    \langle \tilde{\textbf{v}} \rangle = \alpha_c (\xi \delta + r_c) (1-A(\xi)), \\ \\
    \langle  \tilde{\nabla} \cdot \tilde{\textbf{v}}(\xi) \rangle = \alpha_c \biggl[ 3 (B(\xi)-A(\xi)) - \dfrac{(\xi \delta + r_c )}{\delta(\epsilon)} \dfrac{\text{d} A}{\text{d} \xi} \biggr].
\end{cases}
\end{equation}
Clearly $\langle \tilde{\textbf{v}} \rangle $ becomes a finite value as $\delta \to 0$, so its gauge function will be a finite number, which is represented as $O(1)$.\\

To establish the gauge function related to $\langle  \tilde{\nabla} \cdot \tilde{\textbf{v}}(\xi) \rangle$ we need to describe $\text{d}A/ \text{d}\xi$, which can be written as:
\begin{multline}
    \dfrac{\text{d} A}{\text{d} \xi} = \delta \biggl(\dfrac{-104.179 + 142.6(\text{r}_c + \delta \xi) - 28.8587(\text{r}_c + \delta \xi)^2}{(\text{r}_c + \delta \xi)^9} +\\- \dfrac{17.6964}{(\text{r}_c + \delta \xi)^6} \biggr) = \delta A_r(\xi).
\end{multline}
From inspection it is evident that $A_r$ is bounded as $\epsilon \to 0$, so $A_r$ = O(1) and, as a consequence, $\text{d} A / \text{d}\xi$ = O($\delta(\epsilon)$) as $\delta \to 0$. Therefore, we can see that $\langle \tilde{\nabla} \cdot \tilde{\textbf{v}}(\xi) \rangle = O(1)$ as $\epsilon \to 0$.

In order to proceed with the calculations which involve the interaction potential, we need to find the gauge functions relative to its first and second order derivatives. After having applied the inner transformation we obtain:
\begin{equation}
    \tilde{U} = \tilde{U}_{\text{min}} \biggl( (\xi \delta(\epsilon) + \tilde{r}_c)^{-12} - (\xi \delta(\epsilon) + \tilde{r}_c)^{-6} \biggl).
\end{equation}
Now we differentiate the previous equation with respect to $\xi$ obtaining:
\begin{multline}
\begin{cases}
	\dfrac{\text{d} \tilde{U}}{\text{d} \xi} = \tilde{U}_{\text{min}} \delta(\epsilon) \biggl[ \dfrac{-12}{(\xi \delta(\epsilon) + \tilde{r}_c)^{13}} + \dfrac{6}{(\xi \delta(\epsilon) + \tilde{r}_c)^{7}}  \biggr] \\ \\
	\dfrac{\text{d}^2 \tilde{U}}{\text{d} \xi^2} = \tilde{U}_{\text{min}} \delta(\epsilon)^2 \biggl[ \dfrac{156}{(\xi \delta(\epsilon) + \tilde{r}_c)^{14}} - \dfrac{42}{(\xi \delta(\epsilon) + \tilde{r}_c)^{8}}  \biggr].
\end{cases}
\end{multline}
It is clear that the trend of  both  derivatives is strictly related to the width of the boundary layer, so 
\begin{equation}
\begin{cases}
    \dfrac{\text{d}\tilde{U}}{\text{d} \xi} = O(\delta(\epsilon)) \\ \\ 
    \dfrac{\text{d}^2 \tilde{U}}{\text{d} \xi^2} = O(\delta(\epsilon)^2).
\end{cases}
\end{equation}
It is possible to write the structure of the two functions with respect to two terms, namely $W(\xi)$ and $X(\xi)$, which are asymptotically bounded as $\epsilon \to 0$:  
\begin{equation}
    \begin{cases}
    W(\xi) = \tilde{U}_{\text{min}} \biggl[ \dfrac{-12}{(\xi \delta(\epsilon) + \tilde{r}_c)^{13}} + \dfrac{6}{(\xi \delta(\epsilon) + \tilde{r}_c)^{7}}  \biggr] \\ \\
    X(\xi) = \tilde{U}_{\text{min}} \biggl[ \dfrac{156}{(\xi \delta(\epsilon) + \tilde{r}_c)^{14}} - \dfrac{42}{(\xi \delta(\epsilon) + \tilde{r}_c)^{8}}  \biggr].
    \end{cases}
\end{equation}
Finally we analyze the hydrodynamic function $G(\xi)$ and, in particular its derivative $\dfrac{\text{d}G}{\text{d}\xi}$:
\begin{equation}
\begin{cases}
	G(\xi) = \dfrac{24(\xi \delta +\tilde{r}_c-1)^2 + 8(\xi \delta +\tilde{r}_c-1)}{24(\xi \delta +\tilde{r}_c-1)^2 + 26(\xi \delta +\tilde{r}_c-1) + 2} \\ \\ 
	\dfrac{\text{d}G}{\text{d} \xi} = \dfrac{4 \delta(\epsilon) }{11} \biggl( \dfrac{2}{(\xi \delta + \tilde{r}_c)^2} + \dfrac{9}{(\xi \delta + 12 \tilde{r}_c -11)^2} \biggr) = \delta(\epsilon) G_r.
\end{cases}
\label{hydrodynamic_function_analysis}
\end{equation}
From inspection of Eq.(\ref{hydrodynamic_function_analysis}) it can be found that, since $G_r$ is asymptotically bounded as $\epsilon \to 0$, $\dfrac{\text{d} G}{\text{d} \xi}$ = $O(\delta(\epsilon))$.
\section{Inner solution calculations}
In this section, the steps related to the evaluation of $g_1^{\text{in}}(\xi)$ are reported:
to solve Eq.(\ref{first_order_inner_equation}) it is necessary to introduce a change of variable
\begin{equation}
    p_1 = \dfrac{\text{d} g_{1}^{\text{in}}}{\text{d} \xi},
\end{equation}
so Eq.(\ref{first_order_inner_equation}) becomes:
\begin{multline}
    \dfrac{\text{d} p_{1}}{\text{d} \xi} - \dfrac{4 \langle \tilde{\textbf{v}}(\xi) \rangle}{G(\xi)} p_1 = - \biggl[ \biggl( \dfrac{2}{(\xi \epsilon + \tilde{r}_c)} + W(\xi) + \dfrac{G_r}{G} \biggr) \dfrac{\text{d} g_{0}^{\text{in}}}{\text{d} \xi} + \\ -\dfrac{ \langle \tilde{\nabla}_{\xi} \cdot \tilde{\textbf{v}}(\xi) \rangle}{G(\xi)} 4 g_0^{\text{in}(\xi)} \biggr] = 0.
\end{multline}
Now, it is possible to treat this equation as a first order ODE which will be solved through the introduction of an integrating factor:
\begin{equation}
    \mu(\xi) = \exp \biggl[ \biggl(\int_0^{\xi} - 4 \dfrac{\langle \tilde{\textbf{v}}(\xi) \rangle}{G(\xi)} d\xi \biggr) \biggr],
\end{equation}
which transforms the above mentioned equation as
\begin{multline}
    \dfrac{\text{d} [\mu(\xi) p_1]}{\text{d} \xi} = - \biggl[ \biggl( \dfrac{2}{(\xi \epsilon + \tilde{r}_c)} + W(\xi) + \dfrac{G_r}{G(\xi)} \biggr) \dfrac{\text{d} g_{0}^{\text{in}}}{\text{d} \xi} + \\ - 4 \dfrac{\langle \tilde{\nabla}_{\xi} \cdot \tilde{\textbf{v}}(\xi) \rangle}{G(\xi)} g_0^{\text{in}}].
\end{multline}
The previous equation can be solved analytically, ending up with
\begin{multline}
    p_1\mu(\xi) = \dfrac{\text{d} g_{1}^{\text{in}}}{\text{d} \xi} \mu(\xi) =\int_0^{\xi}\biggl[ \biggl( \dfrac{2}{(\xi \epsilon + \tilde{r}_c)} + W(\xi) + \dfrac{G_r}{G} \biggr) \dfrac{\text{d} g_{0}^{\text{in}}}{\text{d} \xi} + \\ - 4 \dfrac{\langle \tilde{\nabla}_{\xi} \cdot \tilde{\textbf{v}}(\xi) \rangle}{G(\xi)} g_0^{\text{in}} \biggr] \text{d} \xi.
    \label{first_order_inner_equation_2}
 \end{multline}
 
Finally, it is possible to solve Eq.(\ref{first_order_inner_equation}) obtaining:
\begin{multline}
    g_{1}^{\text{in}} = C_3 + \int_0^{\xi} \biggl \{ C_2 - \int_0^{\xi} \biggl[ \biggl( \dfrac{2}{(\xi \epsilon + \tilde{r}_c)} + W(\xi) + \dfrac{G_r}{G} \biggr) \dfrac{\text{d} g_{0}^{\text{in}}}{\text{d} \xi} + \\ - 4 \dfrac{\langle \tilde{\nabla}_{\xi} \cdot \tilde{\textbf{v}}(\xi) \rangle}{G(\xi)} g_0^{\text{in}} \biggr] \text{d} \xi \biggr \}  \times \\ \times \exp \biggl[ \biggl(\int_0^{\xi} 4 \dfrac{\langle \tilde{\textbf{v}}(\xi) \rangle}{G(\xi)}  d\xi \biggr) \biggr] \text{d} \xi.
\end{multline}

\bibliography{bibFreons}

\begin{thebibliography}{99}

\bibitem{Larsen}
J. Larsen, \textit{Foundations of High-energy-density Physics: Physical Processes of Matter at Extreme Conditions}, (Cambridge University Press, 2017).

\bibitem{Hansen}
J.P. Hansen, \textit{Statistical mechanics of dense ionized matter. I. Equilibrium properties of the classical one-component plasma} Physical Review A, \textbf{8}, 3096-3109 (1973).

\bibitem{BatchelorGreen}
G.K. Batchelor, J.T. Green, \textit{The hydrodynamic interaction of two small freely-moving spheres in a linear flow field},Journal of Fluid Mechanics \textbf{56},375-400(1972).

\bibitem{Honig}
E.P. Honig, G.J.Roebersen, P.H. Wieresema, \textit{Effect of hydrodynamic interaction on the coagulation rate of hydrophobic colloids}, J. Coll. Interface Sci. 36, 97 (1971).

\bibitem{VandeVen}
T.G.M. Van de Ven, S.G. Mason, \textit{The microrehology of colloidal dispersions. VIII. Effect of shear on perikinetic doublet formation.}, Colloid \& Polymer Science, \textbf{255}, 794-804 (1977).

\bibitem{Morris}
J. F. Morris, B. Katyal, \textit{Microstructure from simulated Brownian suspension flows at large shear rate}, Physics of Fluid \textbf{14}, 1920-1937 (2002).

\bibitem{Smoluchowski}
M. Smoluchowski,\textit{Versuch einer Mathematischen Theorie der Koagulationskinetik Kolloider Loesungen}, Z. Phys. Chem \textbf{92} 120-168 (1917).	

\bibitem{Ronis}
D. Ronis, \textit{Theory of fluctuations in colloidal suspensions undergoing steady shear flow}, Phys. Rev. A 29, 1453 (1984).

\bibitem{Hess}
J.F. Schwarzl and S. Hess, \textit{Shear-flow-induced distortion of the structure of a fluid: Application of a simple kinetic equation}, Phys. Rev. A 33, 4277 (1986).

\bibitem{Dhont}
J.K.G. Dhont, \textit{On the distortion of the static structure factor of colloidal fluids in shear flow}, Journal of Fluid Mechanics \textbf{204}, 421-431 (1989).

\bibitem{Blaw}
J. Blawzdziewicz and G. Szamel, \textit{Structure and rheology of semidilute suspension in shear flow}, Phys. Rev. E 48, 4632 (1993).

\bibitem{Ackerson}
N. A. Clark and B.J. Ackerson, \textit{Observation of the Coupling of Concentration Fluctuations to Steady-State Shear Flow}, Phys. Rev. Lett. 44, 1005 (1980). 

\bibitem{Clark}
B.J. Ackerson and N.A. Clark, \textit{Sheared colloidal suspensions}, Physica A 118, 221 (1983). 

\bibitem{Brader}
J. M. Brader, M. E. Cates, M. Fuchs, \textit{First-Principles Constitutive Equation for Suspension Rheology}, Phys. Rev. Lett. 101, 138301 (2008).

\bibitem{Cates}
M. Fuchs and M.E. Cates, \textit{A mode coupling theory for Brownian particles in homogeneous steady shear flows}, J. Rheol. 53, 957-1000 (2009).

\bibitem{Onuki}
A. Onuki,
Phase Transition Dynamics (Cambridge University Press, Cambridge, 2004).

\bibitem{Feke}
D.L. Feke, W.R. Schowalter, \textit{The effect of Brownian diffusion on shear-induced coagulation of colloidal dispersions},
Journal of Fluid Mechanics,\textbf{133}, 17-35 (1983).

\bibitem{Batchelor}
G. K. Batchelor, J. T. Green, \textit{The determination of the bulk stress in a suspension of spherical particles to order} $c^2$, Journal of Fluid Mechanics \textbf{56}, 401-427 (1972).

\bibitem{MorrisBrady}
J. F. Brady, J. F. Morris, \textit{Microstructure of strongly sheared suspensions and its impact on rheology and diffusion}, Journal of Fluid Mechanics \textbf{348}, 103-139 (1997).

\bibitem{ZacconeNess}
C. Ness, A. Zaccone, \textit{Effect of hydrodynamic interactions on the lifetime of colloidal bonds}, Industrial \& Engineering Chemical Research, \textbf{56}, 3726-3732 (2017). 

\bibitem{BenderOrszag}
C. M. Bender, S.A. Orszag, \textit{Advanced mathematical methods for scientists and engineers I: Asymptotic methods and perturbation theory} (Springer Science \& Business Media, New York, 1999).

\bibitem{ZacconePRE2009}
A.Zaccone, H. Wu, D.Gentili, M.Morbidelli,
\textit{Theory of activated process under shear with application to shear-induced aggregation of colloids},
Physical Review E \textbf{80}, 051404 (2009).

\bibitem{NazockdastMorris}
E.Nazockdast, J.F. Morris,
\textit{Microstructural theory and the rheology of concentrated colloidal suspensions},
Journal of Fluid Mechanics \textbf{713}, 420-452 (2012).

\bibitem{Hinch}
J.Hinch, \textit{Perturbation methods}, (Cambridge University Press, 1991).

\bibitem{VanDyke}
M. Van Dyke, \textit{Perturbation methods in fluid mechanics}, (Parabolic Press, Stanford, 1975).

\bibitem{shear_flow_spherical_coordinates}
P.M. Adler, \textit{Interaction of unequal spheres. I. Hydrodynamic interaction: colloidal forces}, Journal of colloidal and interface science \textbf{84}, 461-473 (1981).

\bibitem{Yurchenko}
S.O. Yurchenko, \textit{The shortest-graph method for calculation of the pair correlation function in crystalline systems}, The Journal of Chemical Physics 140, 134502 (2014).

\bibitem{Glienke}
M.L. Larsen, R.A. Shaw, A.B. Kostinski, S. Glienke \textit{Fine-scale droplet clustering in atmospheric clouds: 3D Radial Distribution Function from airborne digital photography}, Phys. Rev. Lett. \textbf{121}, 204501 (2018).

\bibitem{Mura} F. Mura and A. Zaccone, \textit{Effects of shear flow on phase nucleation and crystallization}, Phys. Rev. E 93, 042803 (2016).

\bibitem{Melis}
S. Melis, M. Verduyn, G. Storti, M. Morbidelli, J. Baldyga, \textit{Effect of fluid motion on the aggregation of small particles subject to interaction forces}, AlChe Journal \textbf{45}, 1383-1393 (1999).

\end{thebibliography}
\bibliographystyle{plain}

\end{document}